\documentclass[aps, pra,
 preprint,
 groupedaddress, amsfonts,
               amsmath, amssymb, showpacs, nofootinbib]{revtex4-1}
\usepackage{microtype}
\usepackage{graphicx}
\usepackage{epstopdf}
\usepackage[utf8]{inputenc}
\usepackage[T1]{fontenc}
\usepackage[usenames,dvipsnames]{xcolor}
\usepackage{hyperref}
\usepackage{amsthm}

\newcommand{\apm}{{a_{\pm}}}
\newcommand{\eps}{{\epsilon}}
\newcommand{\epsp}{{\epsilon_{+}}}
\newcommand{\epsm}{{\epsilon_{-}}}
\newcommand{\epspm}{{\epsilon_{\pm}}}
\newcommand{\mass}{{m_{e}}} 
\newcommand{\cH}{{\mathcal{H}}} 
\newcommand{\cF}{{\mathcal{F}}} 
\newcommand{\xip}{{\xi_{+}}} 
\newcommand{\xipmax}{{\xi_{+}^{\max}}} 
\newcommand{\xipmin}{{\xi_{+}^{\min}}} 
\newcommand{\xim}{{\xi_{-}}} 
\newcommand{\ximmax}{{\xi_{-}^{\max}}} 
\newcommand{\ximmin}{{\xi_{-}^{\min}}} 
\newcommand{\xipm}{{\xi_{\pm}}} 
\newcommand{\xipmmin}{{\xi_{\pm}^{\min}}} 
\newcommand{\xipmmax}{{\xi_{\pm}^{\max}}} 
\newcommand{\xiz}{{\xi_{0}}} 
\newcommand{\pip}{{\pi_{+}}} 
\newcommand{\pim}{{\pi_{-}}} 
\newcommand{\pipm}{{\pi_{\pm}}} 
\newcommand{\piz}{{\pi_{0}}} 
\newcommand{\thp}{{\theta_{+}}} 
\newcommand{\thpz}{{\theta_{+}^{(0)}}} 
\newcommand{\thm}{{\theta_{-}}} 
\newcommand{\thmz}{{\theta_{-}^{(0)}}} 
\newcommand{\thpm}{{\theta_{\pm}}} 
\newcommand{\thz}{{\theta_{0}}} 
\newcommand{\thzz}{{\theta_{0}^{(0)}}} 
\newcommand{\ip}{{I_{+}}} 
\newcommand{\im}{{I_{-}}} 
\newcommand{\ipm}{{I_{\pm}}} 
\newcommand{\iz}{{I_{0}}}
\newcommand{\Ap}{{a_{+}}}
\newcommand{\am}{{a_{-}}}
\newcommand{\psip}{{\psi_{+}}}
\newcommand{\psim}{{\psi_{-}}}
\newcommand{\psipm}{{\psi_{\pm}}}
\newcommand{\chip}{{\chi_{+}}}
\newcommand{\chim}{{\chi_{-}}}
\newcommand{\chipm}{{\chi_{\pm}}}

\newcommand{\ompm}{{\omega_{\pm}}}
\newcommand{\omz}{{\omega_{0}}}
\newcommand{\sgn}{{{\rm sgn}}} 
\newcommand{\cp}{{c_{+}}}
\newcommand{\cm}{{c_{-}}}

\usepackage{color}% \textcolor{#1}{#2}

\begin{document}
\title{Classical calculation of radiative decay rates of hydrogenic Stark states}

\author{Michael Horbatsch}
\affiliation{Department of Physics and Astronomy, York University, Toronto, Ontario M3J 1P3, Canada}

\author{Marko Horbatsch}
\email[]{marko@yorku.ca}
\affiliation{Department of Physics and Astronomy, York University, Toronto, Ontario M3J 1P3, Canada}
\date{\today}
\begin{abstract}
The Kepler-Coulomb problem is solved in parabolic coordinates and the Larmor radiation problem is analyzed to complement a previous study
performed for the usual representation in spherical polar coordinates. 
The problem is then extended to include a weak electric field and a solution in terms of action-angle variables is provided.
A comparison with quantum spontaneous decay rates shows that for
azimuthal quantum number $m=0$ states only transitions to nearby $n-\Delta n$ principal quantum number
states are described properly by the Wentzel-Kramers-Brillouin quantized classical
motions, but that for $m>0$ reasonable results emerge for many values of $\Delta n$. A simple approximate expression for the lifetime of 
$m \ne 0$ states emerges from the semi-classical analysis.
\end{abstract}
%

%\pacs{34.10.+x, 34.50.Gb, 34.70.+e, 36.40.-c}
%

\maketitle
\section{Introduction}
\label{intro}
There has been growing interest in the problem of Rydberg-Stark states, e.g., in the context of the search for
molecules with a large permanent electric dipole moment~\cite{PhysRevLett.85.2458,Li1110}.
For hydrogen-like systems the question of radiative lifetimes of such states  arises  in the context of guiding and trapping highly excited states of atoms and
molecules~\cite{Seiler_2016,Hogan_2016}. For a general review of the properties of such states we refer to Ref.~\cite{Gallagher_1988}.

The semi-classical nature of hydrogenic states with high principal quantum number $n$ has been an ongoing subject of interest~\cite{10.2307/24964422}.
The  Wentzel-Kramers-Brillouin (WKB) treatment of the problem of Stark energy levels has led to an understanding of the field-induced auto-ionization of these states that is
complementary to the quantum mechanical treatment~\cite{BetheSalpeter,LandauLifshits,PhysRevA.24.2491,PhysRevA.26.1775,PhysRevLett.49.867}.
The Stark problem was historically a major driving force for the development of quantum mechanics, for a detailed and technical review of this subject area
we refer to Ref.~\cite{duncan2014trouble}.

Rydberg-Stark states have been investigated in experiments using laser spectroscopy of alkali atoms~\cite{PhysRevLett.36.788,PhysRevA.14.1614,PhysRevLett.41.1463,PhysRevA.20.2251,PhysRevLett.47.83,PhysRevA.14.1614}, and also of atomic hydrogen~\cite{Delsart_1987,PhysRevA.46.5836,PhysRevA.31.530,StarkEng}. 
Ref.~\cite{Delsart_1987} gives a brief review of the early work from the 1930s, both experimental and theoretical, which led to the understanding of the
auto-ionizing character of the Stark states, and the superiority of the wave mechanics approach in the case of stronger electric fields.

The more recent laser-spectroscopy based hydrogen experimental results were found to be in good agreement 
with detailed numerical  Schr\"odinger solutions for the auto-ionizing states~\cite{Luc_Koenig_1980, Luc_Koenig_1980b}.
In order to resolve the parabolic states one applies strong electric fields and this leads to Stark mixing of levels with different principal quantum number $n$. One also has to take
into account the fine structure of the lower levels. For the present work on radiative decays we will limit ourselves to weak electric fields, such that auto-ionization and level mixing is suppressed. As explained in Ref.~\cite{Luc_Koenig_1980} one can define an energy criterion on the basis of the location of the saddle in the combined Coulomb and electric field potential energies. Fig.~5  in Ref.~\cite{Luc_Koenig_1980} demonstrates both the range of electric fields to avoid $n$-mixing, and to remain in the tunnelling regime, which
is a requirement for the applicability of a classical analysis of radiative decays.

The problem of radiative decays can be addressed in the semi-classical theory making use of the Fourier series representation of Kepler-type orbits and
the Larmor formula~\cite{LandauLifshitsEM,Jackson}. The so-called Heisenberg correspondence works for transitions to the nearest-neighbor states only, i.e., $\Delta n=1$ transitions.
For higher-$\Delta n$ transitions the equidistant in energy Fourier terms need to be collected in some appropriate way. This was accomplished in Ref.~\cite{PhysRevA.71.020501},
where it was also shown that the approach leads to acceptable transition rates for angular-momentum-decreasing transitions ($\Delta l = -1$), and to a lesser
degree for angular-momentum-increasing transitions ($\Delta l = +1$) when the decaying state was not an s-state (i.e., $l>0$). The agreement with the quantum
transition rates was shown to improve with increasing $l$-value of the state considered, particularly for higher principal quantum number $n$.
This work was then extended to deal with the lifetimes of diamagnetic states~\cite{PhysRevA.72.033405} on the basis of their semi-classical description~\cite{PhysRevA.28.7}.
Since the hydrogenic radiative decay treatment excluded $l=0$ states the problem of the magnetic-field states had to be restricted to non-zero azimuthal (magnetic)
quantum number $m$.

For the Stark problem a similar strategy was then adopted to deal with $m \ne 0$ states. A simple lifetime formula was derived which demonstrated that for theses states
the decay rate would depend only on $n$ and $m$, and not on the parabolic quantum numbers $n_1$ and $n_2$~\cite{2006APS..DMP.G1021H}.
In the present work we include the treatment of $m=0$ states for which one finds a strong dependence on $\mu=n_1-n_2$ for the lifetimes. The semi-classical
treatment works well for many of the states, but some exceptions do occur. While with present-day computers there is no practical need to avoid quantum rate
calculations, we do present our findings mostly to highlight to which extent classical-quantum correspondence can work.
Thus, the work aims to illustrate the level of sophistication in classical mechanics and electromagnetism that was developed before quantum theory
was established. 

Compared to the previously treated problem of hydrogenic angular momentum eigenstates one has to step up the effort in Hamilton-Jacobi theory.
First one addresses the problem of field-free Coulomb-Stark states which can be described by ensembles of elliptic orbits with well-defined eccentricity range. 
Once the weak electric field is turned on, the orbits are no longer closed, and are dealt with at the level of classical perturbation theory, which is justified in the weak-field limit. As an aside we note that the Kepler-Stark problem
is of interest in astrophysics, and an exact solution without recourse to perturbation theory can be obtained~\cite{exactStark}.

The present work deals with electric dipole (E1) transitions and compares non-relativistic Schr\"odinger results in the dipole approximation with corresponding results
from the semi-classical Kepler problem with Larmor radiation. One may be concerned that for transitions corresponding to large changes in principal quantum number
the emitted radiation has a wavelength that is no longer considerably larger than the size of the original orbit, and thus might invalidate the use of the dipole approximation~\cite{Corney}.
Tables of quantum-mechanical electric quadrupole (E2) transition rates are available for atomic hydrogen fine-structure  states with principal quantum number up to $n=25$~\cite{Jitrik_2004}. We have verified that the E1 transition rates calculated for $n=20$ Stark states exceed the E2 rates by at least two orders of magnitude.
As an aside we note that the classical quadrupole radiation problem plays an important role in gravitational physics~\cite{PhysRev.131.435}, and that this work can be adapted to the Kepler-Coulomb problem.

The paper is organized as follows. In Sect.~\ref{sec:theory1} we introduce the theoretical method for the current work, namely the semi-classical approach.
 In Sect.~\ref{sec:theory2} we review how the quantum spontaneous decay rates for Coulomb-Stark states are obtained from the field-free hydrogenic states
by combining angular momentum contributions.
In Sect.~\ref{sec:res} we present our numerical results and we comment on the classical-quantum correspondence for the parabolic oscillators which form the
basis of the Coulomb-Stark states. 
The paper ends with a few concluding remarks in Sect.~\ref{sec:conclusions}.
%Atomic units, characterized by $\hbar=m_e=e=4\pi\epsilon_0=1$, are used unless otherwise stated.

\section{Theory}
\label{sec:model}

\subsection{Semi-classical approach}
\label{sec:theory1}

\subsubsection{Equations of motion and conserved observables}
\label{sec:theory1sub1}

A review of the mathematical aspects of the quantum and classical  hydrogen atom problem in different coordinate systems can be found in the book by Cordani~\cite{Cordani}.
The quantum mechanical solutions of the hydrogen atom in parabolic coordinates have been described in detail in Ref.~\cite{Hey_2007}.
We use the parabolic coordinates defined as
\begin{equation}
\label{eq:eq1}
\xi_\pm = \sqrt{x^2+y^2+z^2} \pm z \  , \quad \xi_0=\arctan{(y/x)} \ .
\end{equation}
The Schr\"odinger problem separates in these coordinates and the eigenfunctions can be expressed as products of parabolic oscillator solutions times the usual
azimuthal solution, i.e., writing in the usual form $\xi_0=\varphi$ the quantum solutions are
\begin{equation}
\label{eq:eq2}
\psi(\xi_{+}, \xi_{-}, \varphi) = f_+(\xi_{+}) f_-(\xi_{-}) e^{i m \varphi} \ .
\end{equation}
For a detailed description of the product-form wavefunctions see Ref.~\cite{Hey_2007}.
We note that the two parabolic oscillator solutions are independent of each other. Using the semi-classical WKB approach the radial equations can be re-written in such a form that the parabolic oscillator actions are quantized as half-integers,
 $n_1+\tfrac{1}{2}$ and $n_2+\tfrac{1}{2}$ respectively~\cite{PhysRevA.26.1775,duncan2014trouble}.

We are now taking the classical problem that emerges in the semi-classical quantization approach and pursue the calculation of Larmor radiation.
%The canonical momenta associated with the parabolic coordinates can be expressed in terms of the cartesian components as
%\begin{eqnarray}
%\pipm &=& \pm \frac{1}{2} \left[
%p_{z} + \frac{xp_{x} + yp_{y}}{z \pm \sqrt{x^2+y^2+z^2}}
%\right] \,, \\
%\piz &=& xp_{y}-yp_{x} \,.
%\end{eqnarray}
In order to solve the problem analytically, carry out a canonical transformation
on the six-dimensional Hamiltonian phase space from Cartesian coordinates and momenta 
$(x,y,z,p_{x},p_{y},p_{z})$ to the parabolic coordinates and momenta
$(\xip,\xim,\xiz,\pip,\pim,\piz)$, given by equation~(\ref{eq:eq1}), and
\begin{eqnarray}
\label{eq:cart2par_1}
%\xipm &=& \sqrt{x^{2}+y^{2}+z^{2}} \pm z \,, \\
%\xiz &=& \arctan{\frac{y}{x}} \,, \\
\pipm &=& \pm \frac{1}{2} \left[
p_{z} + \frac{xp_{x} + yp_{y}}{z \pm \sqrt{x^2+y^2+z^2}}
\right] \,, \\
\piz &=& xp_{y}-yp_{x} \,.
\end{eqnarray}

The inverse transformation is given by
\begin{eqnarray}
x &=& \sqrt{\xip\xim}\cos \xiz \,,
\\
y &=& \sqrt{\xip\xim}\sin \xiz \,,
\\
z &=& \frac{\xip-\xim}{2} \,,
\\
p_x &=&  2 \left( \frac{\pip+\pim}{\xip+\xim}\right) \sqrt{\xip\xim}\cos \xiz
 - \frac{\piz}{\sqrt{\xip\xim}} \sin \xiz  \,,
\\
p_y &=&  2 \left( \frac{\pip+\pim}{\xip+\xim}\right) \sqrt{\xip\xim}\sin \xiz
 + \frac{\piz}{\sqrt{\xip\xim}} \cos \xiz  \,,
\\
p_z &=& 2 \, \frac{\xip\pip - \xim\pim}{\xip + \xim} \,.
\end{eqnarray}

The variables $\xipm$ are both restricted to be non-negative.
In the three-dimensional space through which the electron moves, the 
surface $\xip = C > 0$ is a two-dimensional paraboloid which has
its vertex on the positive $z$ axis, and which opens downwards into
the negative $z$ direction:
\begin{equation}
\label{eq:paraboloid}
z = \frac{C}{2} - \frac{x^2+y^2}{2C} \,.
\end{equation}
The surface $\xip=0$ is the one-dimensional non-positive $z$ axis.
Likewise, the surface $\xim=C > 0$ is a two-dimensional paraboloid
which has its vertex on the negative $z$ axis, and which opens upwards
into the positive $z$ direction:
\begin{equation}
z = - \frac{C}{2} + \frac{x^2+y^2}{2C} \,.
\end{equation}
The surface $\xim=0$ is the one-dimensional non-negative $z$ axis.
The variable $\xiz$ is an azimuthal angle, and is restricted to be 
in the range $0 \leq \xiz < 2\pi$. It is undefined on the $z$ axis.
In terms of the parabolic coordinates $\xipm$, the positive $z$ axis corresponds
to $(\xip=2z,\,\xim=0)$, the negative $z$ axis corresponds to
$(\xip=0,\,\xim=-2z)$, and the origin corresponds to 
$(\xip=0,\,\xim=0)$.
More generally, the region $z>0$ corresponds to $\xip > \xim$, 
the region $z<0$ corresponds to $\xip < \xim$, and the plane
$z=0$ corresponds to $\xip = \xim$.
The variables $\pipm$ are linear momenta, whose values are 
not subject to any restrictions. The variable $\piz=L_z$ is the $z$-component of 
angular momentum, also not subject to any restrictions.

In terms of the parabolic coordinates and momenta, the Hamiltonian is given by
\begin{eqnarray}
\label{eq:ham_par}
\cH(\xip,\xim,\xiz,\pip,\pim,\piz) &=&
\frac{2}{\mass} \frac{\xip\pip^{2} + \xim\pim^{2}}{\xip + \xim} 
+ \frac{\piz^{2}}{2\mass\xip\xim}
\nonumber
\\
&&
- \frac{2k}{\xip + \xim} + \frac{F}{2}\left( \xip - \xim\right) \, .
\end{eqnarray}
Here $\mass$ is the electron mass, and the external field ${\vec E}$ is chosen along the positive $z$-direction.
The strength parameters for the Coulomb attraction  and for the electric field are defined as
\begin{equation}
k = \frac{Ze^{2}}{4\pi\epsilon_{0}} = Z \alpha \hbar c \,,
\qquad
F = e |{\vec E}| \,,
\end{equation}
respectively and are given in SI units. Here $\alpha$ is the fine-structure constant. The external field is considered to be a small perturbation, in the sense that 
\begin{equation}
F \ll k/a^2 \,,
\end{equation}
where $a$ is a length scale of the orbit, i.e., the order of magnitude
of the electron-nucleus separation. It can be taken to be the semi-major axis $a$, which is defined in equation~(\ref{eq:kepler_spherical}) below. 

The equations of motion are obtained from
\begin{equation}
\label{eq:ham_xipd}
\dot{\xip} = \frac{\partial \cH}{\partial \pip}
= \frac{4\xip\pip}{\mass(\xip+\xim)} \,,
\end{equation}
\begin{equation}
\label{eq:ham_ximd}
\dot{\xim} = \frac{\partial \cH}{\partial \pim}
= \frac{4\xim\pim}{\mass(\xip + \xim)}\,,
\end{equation}
\begin{equation}
\dot{\xiz} = \frac{\partial \cH}{\partial \piz}
= \frac{\piz}{\mass\xip\xim} \,,
\end{equation}
\begin{equation}
\dot{\pip} = - \frac{\partial \cH}{\partial \xip}
= - \frac{2\xim\left(\pip^2 - \pim^2 \right)}{\mass(\xip+\xim)^2} 
+ \frac{\piz^2}{2\mass\xip^2\xim} - \frac{2k}{(\xip + \xim)^2} - \frac{F}{2} \,,
\end{equation}
\begin{equation}
\dot{\pim} = - \frac{\partial \cH}{\partial \xim}
= \frac{2\xip\left(\pip^2 - \pim^2 \right)}{\mass(\xip + \xim)^2}  
+ \frac{\piz^2}{2\mass\xip\xim^2} 
- \frac{2k}{(\xip + \xim)^2} + \frac{F}{2} \,,
\end{equation}
\begin{equation}
\dot{\piz} = - \frac{\partial \cH}{\partial \xiz} = 0 \,.
\end{equation}

The following remarks can be made about these equations. First of all, it is obvious that the $z$-component of angular momentum is conserved,
i.e., $\piz = L_z =  m \hbar$ is a constant of the motion (even with non-zero electric field, i.e., $F \ne 0$). What is less obvious is that the $z$-component
of the generalized Laplace-Runge-Lenz vector 
\begin{equation}
\vec{A} = \vec{p} \times \vec{L} - \mass k \hat{r} + \frac{\mass F}{2} \left[ \vec{r} \times (\vec{r} \times \hat{z}) \right]
= A_{x} \hat{x} + A_{y} \hat{y} + A_{z} \hat{z}\,,
\end{equation}
%\begin{equation}
%\vec{A} = \vec{p} \times \vec{L} - \mass k \hat{r} 
%= A_{x} \hat{x} + A_{y} \hat{y} + A_{z} \hat{z}\,,
%\end{equation}
%
does remain conserved for $F \ne 0$. In parabolic coordinates, the conserved quantity  becomes
\begin{equation}
\label{eq:az_par}
A_{z} = -\frac{2\xip\xim(\pip^2 - \pim^2)}{\xip + \xim}
+ \frac{\piz^2(\xip - \xim)}{2 \xip \xim} 
-\mass k \, \frac{\xip - \xim}{\xip + \xim} -\frac{\mass F}{2} \xip \xim \,.
\end{equation}

The other Cartesian components $L_x, L_y, A_x, A_y$ are not conserved (even for
weak fields when $F\approx 0$). When one constructs orbits from the parabolic oscillator solutions, one finds precessing ellipses for which
energy conservation implies a constant semi-major
axis value of $a$, which is related to the principal quantum number $n$. The non-conservation of $L_x, L_y$ implies that a range of eccentricities is allowed, and that
the magnitude of orbital angular momentum varies during the orbit.

An interesting difference between the classical and quantum parabolic oscillator problems is that at the level of equations of motion the
two oscillators remain coupled, i.e., they cannot be solved separately for $\xip(t)$ and $\xim(t)$. This is in contrast with the Schr\"odinger solutions
which are products of parabolic oscillator states, cf. equation (\ref{eq:eq2}).

\subsubsection{Action-angle variables and phase curves}
\label{sec:theory1sub2}
Now we construct action-angle variables
$(\thp,\thm,\thz,\ip,\im,\iz)$. The angle variables 
$(\thp,\thm,\thz)$ will be our new coordinates, and the 
action variables $(\ip,\im,\iz)$ will be our new momenta.
The canonical transformation
can be derived from 
\begin{equation}
\pip\dot{\xip} +\pim\dot{\xim} + \piz\dot{\xiz} 
= \ip\dot{\thp} + \im\dot{\thm} +\iz\dot{\thz} + \frac{d\cF}{dt} \,,
\end{equation}
where
\begin{equation}
\mathcal{F} = W - \thp\ip - \thm\im - \thz\iz \,,
\end{equation}
and where $W$ is Hamilton's characteristic function, that depends only on 
the old coordinates $(\xip,\xim,\xiz)$ and new momenta $(\ip,\im,\iz)$. It follows from the above two equations
that 
\begin{equation}
\pip = \frac{\partial W}{\partial \xip} \,,
\qquad
\pim = \frac{\partial W}{\partial \xim} \,,
\qquad
\piz = \frac{\partial W}{\partial \xiz} \,,
\end{equation}
\begin{equation}
\label{eq:ang_vars}
\thp = \frac{\partial W}{\partial \ip} \,,
\qquad
\thm = \frac{\partial W}{\partial \im} \,,
\qquad
\thz = \frac{\partial W}{\partial \iz} \,.
\end{equation}
To obtain concrete formulas for the new coordinates and momenta, we begin by 
finding Hamilton's characteristic function $W$ from the 
time-independent Hamilton-Jacobi equation~\cite{Goldstein}
\begin{equation}
\cH \left(\xip,\xim,\xiz,
\frac{\partial W}{\partial \xip},\frac{\partial W}{\partial \xim},\frac{\partial W}{\partial \xiz}\right) = E \,,
\end{equation}
which, by equation (\ref{eq:ham_par}), reads
\begin{eqnarray}
E &=& \frac{1}{\xip+\xim} \left[ 
\frac{2\xip}{\mass} \left( \frac{\partial W}{\partial \xip}\right)^{2} - k + \frac{F}{2}\xip^2
\right]
\nonumber
\\
&&
+\frac{1}{\xip + \xim}\left[ 
\frac{2\xim}{\mass} \left( \frac{\partial W}{\partial \xim}\right)^{2} - k - \frac{F}{2} \xim^{2}
\right]
\nonumber
\\
&&
+\frac{1}{2\mass\xip\xim} \left( \frac{\partial W}{\partial \xiz} \right)^{2} \,.
\label{eq:hamjac1}
\end{eqnarray}
We will solve this by separation of variables. Begin with the ansatz 
\begin{eqnarray}
\label{eq:sep_ansatz}
W(\xip,\xim,\xiz,\ip,\im,\iz) &=& W_{+}(\xip,\ip,\im,\iz)
\nonumber
\\
&& + W_{-}(\xim,\ip,\im,\iz)
\nonumber
\\
&& + W_{0}(\xiz,\ip,\im,\iz) \,.
\end{eqnarray}
Then equation (\ref{eq:hamjac1}) becomes 
\begin{eqnarray}
E &=& \frac{1}{\xip+\xim} \left[ 
\frac{2\xip}{\mass} \left( \frac{\partial W_{+}}{\partial \xip}\right)^{2} - k + \frac{F}{2}\xip^2
\right]
\nonumber
\\
&&
+\frac{1}{\xip + \xim}\left[ 
\frac{2\xim}{\mass} \left( \frac{\partial W_{-}}{\partial \xim}\right)^{2} - k - \frac{F}{2} \xim^{2}
\right]
\nonumber
\\
&&
+\frac{1}{2\mass\xip\xim} \left( \frac{\partial W_{0}}{\partial \xiz} \right)^{2} \,.
\label{eq:hamjac2}
\end{eqnarray}
Since $\partial W_{0} / \partial \xiz = \piz = L_{z}$ is constant, we can immediately
write down 
\begin{equation}
\label{eq:hamfunc_azimuth}
W_{0} = L_{z} \xiz \,, 
\qquad
\left( \frac{\partial W_{0}}{\partial \xiz} \right)^{2} = L_{z}^{2} \,.
\end{equation}
It now follows from equations (\ref{eq:hamjac2})-(\ref{eq:hamfunc_azimuth})
that we can introduce a separation constant $C_{2}$ by
\begin{eqnarray}
C_{2} &=&
\frac{2\xip}{\mass} \left(\frac{\partial W_{+}}{\partial \xip} \right)^{2} 
+ \frac{L_{z}^{2}}{2 \mass \xip} - E \xip -k + \frac{F}{2} \xip^{2}
\nonumber
\\
&=&
-\frac{2\xim}{\mass} \left(\frac{\partial W_{-}}{\partial \xim} \right)^{2} 
- \frac{L_{z}^{2}}{2 \mass \xim} + E \xim + k + \frac{F}{2} \xim^{2} \,.
\label{eq:hamjac_sep}
\end{eqnarray}
Using equation (\ref{eq:az_par}), we 
can identify this separation constant as
\begin{equation}
C_{2} = -\frac{A_{z}}{\mass} \,.
\end{equation}
Before going on to obtain the functions $W_{\pm}$, we note that 
equation (\ref{eq:hamjac_sep}) yields the following formulas 
for the phase curves of the parabolic oscillators:
\begin{eqnarray}
\label{eq:phasecurvep}
\frac{2\pip^2}{\mass} + \frac{L_{z}^{2}}{2 \mass \xip^2} - \frac{k}{\xip}
+ \frac{A_{z}}{\mass \xip} + \frac{F \xip}{2} = E \,,
\\
\label{eq:phasecurvem}
\frac{2\pim^2}{\mass} + \frac{L_{z}^{2}}{2 \mass \xim^2} - \frac{k}{\xim}
- \frac{A_{z}}{\mass \xim} - \frac{F \xim}{2} = E \,.
\end{eqnarray}

From these phase curves, we can obtain an intuitive understanding of the motion. 
We begin by calculating turning points. If $F \neq 0$, this 
requires formulas for the roots of a cubic equation, which 
are quite complicated. Therefore, we will now set $F = 0$. 
Furthermore, we restrict attention to bound orbits with $E < 0$. 
We obtain
\begin{equation}
\xip^{\rm max} = \Ap(1+\epsp) \,,
\qquad
\xip^{\rm min} = \Ap(1-\epsp) \,,
\end{equation}
\begin{equation}
\xim^{\rm max} = \am(1+\epsm) \,,
\qquad
\xim^{\rm min} = \am(1-\epsm) \,,
\end{equation}
where
\begin{equation}
\label{eq:apm_epspm}
\apm = \frac{k}{2|E|}(1 \mp \tilde{A}_{z}) \,,
\qquad
\epspm = \sqrt{1 - \frac{2|E|L_{z}^{2}}{\mass k^{2} (1 \mp \tilde{A}_{z})^{2}}} \,.
\end{equation}
Here we have introduced the 
dimensionless generalized Laplace-Runge-Lenz vector 
\begin{equation}
\tilde{\vec{A}} = \frac{\vec{A}}{\mass k} \,,
\end{equation}
which satisfies
\begin{equation}
|\tilde{A}_{z}| \leq 1 \,.
\end{equation}
Several remarks are now in order.
\begin{itemize}
\item First of all, we have not yet said anything about ellipses. 
The $\apm$ are not semi-major axes, and the $\epspm$ are not eccentricities. They are simply
formal quantities that can be defined for any oscillator undergoing librational motion between 
two turning points. Their relation to the geometrical properties of the ellipse traced out 
by the electron will be discussed later in equations~(\ref{eq:semimajor_axis}) and~(\ref{eq:eps_delta}).

\item
Second: although the separation of the Hamilton-Jacobi equation yields
phase curves (\ref{eq:phasecurvep}) and (\ref{eq:phasecurvem}) 
of the $\xip$ and $\xim$ oscillators that are independent of each other, 
it does not follow that the motions in time along these phase curves
can be obtained independently. This is clear from equations (\ref{eq:ham_xipd}) and 
(\ref{eq:ham_ximd}).

\item
Third: if $L_{z}=0$, then $\epspm = 1$, and so $\xipm^{\rm min} = 0$. This implies that 
the electron orbit intersects the $z$ axis. However, it does not necessarily imply that
the electron orbit intersects the origin (i.e. collides with the nucleus). A collision
with the nucleus happens when we simultaneously have $\xip=\xim=0$. In order to determine whether such a 
collision occurs, information about the phase difference between the $\xip$ and $\xim$
oscillators is needed. This will be discussed in more detail later.

\item Fourth: if $\tilde{A}_{z} = 0$, then the $\xip$ and $\xim$ oscillator phase curves 
become identical.

\item Fifth: if $\tilde{A}_{z} = 1$, then it can be shown that 
$\xip^{\rm min} = \xip^{\rm max} = 0$ and $|\vec{L}|=0$, so the $\xip$ coordinate is frozen at $\xip = 0$, and 
the motion is confined to the non-positive $z$ axis. This is a straight-line orbit 
which goes through the nucleus. Similarly, if $\tilde{A}_{z} = -1$, then it can be shown that 
$\xim^{\rm min} = \xim^{\rm max} = 0$ and $|\vec{L}|=0$, so the $\xim$ coordiante is frozen at $\xim = 0$, and the 
motion is confined to the non-negative $z$ axis, a straight-line orbit through the nucleus.
\end{itemize}

\subsubsection{Calculation of Hamilton's characteristic function}
\label{sec:theory1sub3}
Now we finally calculate the contributions $W_{\pm}$ to Hamilton's characteristic function. From (\ref{eq:hamjac_sep}), 
we find (still assuming $F=0$ and $E<0$)
\begin{eqnarray}
\label{eq:hamcharfuncp}
\frac{\partial W_{+}}{\partial \xip} &=& 
\frac{\pm 1}{\xip}\sqrt{\frac{\mass |E|}{2}(\xip - \xip^{\rm min})(\xip^{\rm max} - \xip)} \,,
\\
\label{eq:hamcharfuncm}
\frac{\partial W_{-}}{\partial \xim} &=& 
\frac{\pm 1}{\xim}\sqrt{\frac{\mass |E|}{2}(\xim - \xim^{\rm min})(\xim^{\rm max} - \xim)} \,.
\end{eqnarray}
%
%
%In order to integrate (\ref{eq:hamcharfuncp})-(\ref{eq:hamcharfuncm}), introduce auxiliary variables
%$\psi_{\pm}$ by
%
%\begin{equation}
%\xip = \Ap (1 - \epsp \cos \psip) \,,
%\qquad
%\xim = \am (1 - \epsm \cos \psim) \,.
%\end{equation}
%
%
%Formally, these new variables look very similar to the eccentric anomaly which is used to solve the 
%Kepler problem in spherical polar coordinates. However, unlike the eccentric anomaly,
%they do not have any immediate geometric 
%interpretation. In terms of $\psipm$, equations (\ref{eq:hamcharfuncp})-(\ref{eq:hamcharfuncm}) read
%
%
%
%In addition to yielding simple differential equations (\ref{eq:hamcharfunc_psi}) for the contributions to 
%Hamilton's characteristic function, the variables $\psipm$ have the
%benefit of eliminating the need for the thorny $\pm$ signs in front of the square roots --- a single
%closed phase curve of the $\xip$ oscillator is traced out as $\psip$ goes from $0$ to $2\pi$.

%\end{itemize}
%
In order to solve (\ref{eq:hamcharfuncp})-(\ref{eq:hamcharfuncm}), and also for other future 
calculations, we will need to evaluate 
integrals along phase curves (\ref{eq:phasecurvep}) and (\ref{eq:phasecurvem}) in
the $(\xip,\pip)$ and $(\xim,\pim)$ planes, respectively. 
We will parametrize phase curve (\ref{eq:phasecurvep}) with the auxiliary variable
$\psip$, and we will parametrize phase curve (\ref{eq:phasecurvem}) with the 
auxiliary variable $\psim$. Explicitly, these parametrizations are given by 
\begin{equation}
\label{eq:psip}
\xip = \Ap (1 - \epsp \cos \psip) \,,
\qquad
\pip = \sqrt{\frac{\mass |E|}{2}} \frac{\epsp \sin \psip}{1 - \epsp \cos \psip} \,,
\end{equation}
\begin{equation}
\label{eq:psim}
\xim = \am (1 - \epsm \cos \psim) \,,
\qquad
\pim = \sqrt{\frac{\mass |E|}{2}} \frac{\epsm \sin \psim}{1 - \epsm \cos \psim} \,,
\end{equation}
where the thorny $\pm$ signs appearing in front of the square roots in
(\ref{eq:hamcharfuncp})-(\ref{eq:hamcharfuncm}) have been eliminated --- as $\psipm$ runs
from $-\pi$ to $+\pi$, a full phase curve is traced out in the direction
of phase flow, which is clockwise.

Formally, the variables $\psipm$ look very similar to the eccentric anomaly which is used to solve the 
Kepler problem in spherical polar coordinates. However, unlike the eccentric anomaly,
$\psipm$ do not have any immediate geometric 
interpretation.
We find
\begin{eqnarray}
\label{eq:wp_quad}
W_{+} = \Ap \epsp^2 \sqrt{\frac{\mass |E|}{2}} \int_{0}^{\psip} d\lambda \frac{\sin^2 \lambda}{1 - \epsp \cos \lambda} \,,
\\
\label{eq:wm_quad}
W_{-} = \am \epsm^2 \sqrt{\frac{\mass |E|}{2}} \int_{0}^{\psim} d\lambda \frac{\sin^2 \lambda}{1 - \epsm \cos \lambda} \,.
\end{eqnarray}
Prior to discussing the evaluation of the integral appearing in the above equations, we make two remarks:
\begin{itemize}
\item
Recall from (\ref{eq:sep_ansatz}) that $W_{+}$ is a function of $\xip$, and the three
action variables (which have not yet been 
calculated explicitly). Equation (\ref{eq:wp_quad})
should therefore be understood
in the following sense. Given a value of $\xip$ between $\xip^{\min}$ and $\xip^{\max}$,
the first equation of (\ref{eq:psip}) should be used to calculate the corresponding 
value of $\psip$ that lies between $0$ and $\pi$. This value of $\psip$ should then be substituted into the upper limit
of integration in (\ref{eq:wp_quad}).
In the same way, $W_{-}$ is a function of $\xim$ and the three action variables.

\item
Note that there is nothing special or unique about the lower integration limits
in equations (\ref{eq:wp_quad})-(\ref{eq:wm_quad}). Changing them amounts to 
adding an overall constant to Hamilton's characteristic function, which does not affect the physics, and 
corresponds to multiplying the wave function by an overall phase in quantum mechanics. However, our particular choice
of integration limits, along with our choices for the ranges of $\psipm$, simplify certain branch issues that
arise when dealing with inverse trigonometric functions.
\end{itemize}
Evaluating the integral appearing in equations (\ref{eq:wp_quad})-(\ref{eq:wm_quad}) yields
\begin{equation}
\label{eq:int1}
\int_{0}^{\psi} \frac{\sin^2 \lambda \, d\lambda}{1 - \epsilon \cos \lambda} = 
\frac{1}{\epsilon^2} \left[
\psi + \epsilon \sin \psi -2 \sqrt{1-\epsilon^2}
\arctan \left( \sqrt{\frac{1 + \epsilon}{1 - \epsilon}} \tan \frac{\psi}{2}\right)
\right] \,.
\end{equation}
Two remarks:
\begin{itemize}
\item
The above formula holds for $0 < \epsilon < 1$, and can be extended by
continuity to the cases $\epsilon = 0$ and $\epsilon = 1$. The procedure of integration
is interchangeable with the procedure of taking the limit $\epsilon \to 0$
or $\epsilon \to 1$.

\item
From basic properties of integrals we conclude that the left-hand side of (\ref{eq:int1}) is a continuous function of $\psi$.
Therefore, the right-hand side is also a continuous function of $\psi$. However, in order for this continuity
to be properly realized, the correct branch of the arctangent function must be chosen. Based on earlier discussions,
we only need to evaluate this integral for $\psi$ between $0$ and $\pi$, and therefore, we can simply take
the principal branch of the arctangent function, which maps the real line to the interval $(-\pi/2,\pi/2)$.
\end{itemize}
To simplify the final formulas for $W_{\pm}$ and $\thpm$, we will introduce new quantities $\chipm$, defined by
\begin{equation}
\label{eq:chipm}
\tan \left( \frac{\chip}{2}\right) = 
\sqrt{\frac{1+\epsp}{1-\epsp}} \tan \left(\frac{\psip}{2} \right) \,,
\qquad
\tan \left( \frac{\chim}{2}\right) = 
\sqrt{\frac{1+\epsm}{1-\epsm}} \tan \left(\frac{\psim}{2} \right) \,,
\end{equation}
and the conditions that $\chip=0$ whenever $\psip=0$, and that 
$\chip$ is a continuous function of $\psip$. And similarly for 
$\chim$ and $\psim$. These definitions are only made for 
$0 \leq \epspm < 1$, and the case $\epspm = 1$ will be discussed later.
With these choices, $\chip$ runs from 
$-\pi$ to $+\pi$ as $\psip$ runs over the same range. We can think
of $\chip$ as a new parameter that can be used to parametrize the
phase curves in the $(\xip,\pip)$ plane. 
With the definition of $\chipm$ in hand, 
we can now write relatively simple formulas
for $W_{\pm}$ in terms of $\psipm$ and $\chipm$:
\begin{eqnarray}
\label{eq:wp_fin}
W_{+} = \Ap \sqrt{\frac{\mass |E|}{2}} \left[
\psip + \epsp \sin \psip - \sqrt{1-\epsp^2}  \chip
\right] \,,
\\
\label{eq:wm_fin}
W_{-} = \am \sqrt{\frac{\mass |E|}{2}} \left[
\psim + \epsm \sin \psim - \sqrt{1-\epsm^2}  \chim
\right] \,.
\end{eqnarray}
Before proceeding, we make several remarks:
\begin{itemize}
\item As already discussed, $W_{+}$ is fundamentally a function of the variable $\xip$, and the 
three action variables $\iz,\ipm$ that have not yet been calculated. Given a value of 
$\xip$ between $\xip^{\min}$ and $\xip^{\max}$, the value of $W_{+}$ is calculated as follows. 
The first equation of (\ref{eq:psip}) is used to calculate a unique value of $\psip$ in the range from 
$0$ to $\pi$. Then, this value of $\psip$ is used in the first equation of (\ref{eq:chipm}) to obtain a unique
value of $\chip$ in the range from $0$ to $\pi$. The values of $\psip$ and $\chip$ so obtained 
are to be substituted into equation (\ref{eq:wp_fin}).

\item Equation (\ref{eq:chipm}) has the same form as the relationship between the eccentric and true 
anomalies in the Kepler problem formulated in spherical polar coordinates. (The true anomaly is the angle between
the periastron and the orbiting particle). However, in our problem here, the quantities $\chipm$ are simply 
formal variables that parametrize the phase curves, and have no immediate geometric interpretation as angles.

\item If $\epsp=0$, then $\chip = \psip$. In this case, we have $\xip^{\min} = \xip^{\max}$, and the
$\xip$ coordinate is frozen. The same statement holds for the `minus' variables.

\item In the limit $\epsp \to 1$, the first equation of (\ref{eq:chipm}) becomes $\chip = (\sgn \psip) \cdot \pi$, and 
continuity is lost. However, we do not need the variable $\chip$ when $\epsp=1$, due to the 
factor of $\sqrt{1-\epsilon^2}$ appearing in front of the arctangent in equation (\ref{eq:int1}).
\end{itemize}
Now we have finished computing Hamilton's characteristic function. To put everything together, we need to add
equation (\ref{eq:wp_fin}), equation (\ref{eq:wm_fin}), and the first equation of (\ref{eq:hamfunc_azimuth}).

\subsubsection{Calculation of action variables and semi-classical quantization}
\label{sec:theory1sub4}
Next, we turn to the calculation of the action variables $\iz,\ipm$. 
Note that all the equations that we have written, up until now, do not yet 
tell us how these action variables are defined. Since $\piz$ is constant, we simply
take
\begin{equation}
\iz = \piz = L_{z} \,.
\end{equation}
For $\ipm$, we use the standard prescription
\begin{equation}
\label{eq:action_lineint}
\ipm = \frac{1}{2\pi}\oint \pipm d\xipm \,,
\end{equation}
where the line integrals are to be evaluated along phase curves, in the direction of 
phase flow.
Note that not all authors include the factors of $2\pi$. We include these factors, so that
the semi-classical quantization conditions read
\begin{equation}
\iz = m\hbar \,,
\qquad
\ipm = \left( n_{\pm} + \frac{1}{2} \right) \hbar \,,
\end{equation}
and
the three fundamental {\em angular} frequencies of the multiply-periodic motion are given by 
\begin{equation}
\omz = \frac{\partial \cH}{\partial \iz} \,,
\qquad
\ompm = \frac{\partial \cH}{\partial \ipm} \,.
\end{equation}
Note that  $\omz$ can be of either sign, which is equal to the sign of $L_z$.
Also note that the quantization of $\ipm$ to half-integer multiples of $\hbar$ takes into account the Maslov indices found from the semi-classical WKB treatment
of the Schr\"odinger problem.

To calculate the line integrals in (\ref{eq:action_lineint}), 
we use equations (\ref{eq:psip}) and (\ref{eq:psim}) to parametrize the phase curves, and take the 
limit $\psi \to \pi$ in equation (\ref{eq:int1}) to evaluate the resulting integrals, obtaining 
\begin{eqnarray}
\ip &=& \frac{\Ap \epsp^2}{2\pi}
\sqrt{\frac{\mass |E|}{2}}
\int_{-\pi}^{\pi} \frac{\sin^2 \psip d \psip}{1- \epsp \cos \psip}
\nonumber
\\
&=& \Ap \sqrt{\frac{\mass |E|}{2}} \left( 1 - \sqrt{1-\epsp^2}\right) \,,
\\
\im &=& \frac{\am \epsm^2}{2\pi}
\sqrt{\frac{\mass |E|}{2}}
\int_{-\pi}^{\pi} \frac{\sin^2 \psim d \psim}{1- \epsm \cos \psim}
\nonumber
\\
&=& \am \sqrt{\frac{\mass |E|}{2}} \left( 1 - \sqrt{1-\epsm^2}\right) \,.
\end{eqnarray}
Before turning to the calculation of the angle variables, we express several important quantities 
in terms of the actions. First define the {\em principal action}
\begin{equation}
\label{eq:principalaction}
I = \ip + \im + |\iz|  =  (n_{+} + \tfrac{1}{2}) \hbar + (n_{-} + \tfrac{1}{2}) \hbar  + |m| \hbar = (n_{+} + n_{-} + |m| + 1)\hbar  \,,
\end{equation}
and the {\em electric action}
\begin{equation}
I_{e} = \ip - \im  = (n_{+} - n_{-})\hbar \,.
\end{equation}
After some algebra, we find that
\begin{equation}
\label{eq:ham_az_act}
\cH = E = -\frac{\mass k^2}{2I^2} \,,
\qquad
\tilde{A}_{z} = -\frac{I_{e}}{I} \,,
\end{equation}
and
\begin{eqnarray}
\Ap = \frac{I(I+I_{e})}{\mass k} \,,
\qquad
\am = \frac{I(I-I_{e})}{\mass k} \,,
\\
\epsp = \sqrt{1 - \frac{I_{0}^{2}}{(I+I_{e})^{2}}} \,,
\qquad
\epsm = \sqrt{1 - \frac{I_{0}^{2}}{(I-I_{e})^{2}}} \,.
\end{eqnarray}

\subsubsection{Calculation of angle variables}
\label{sec:theory1sub5}
Next, we calculate the angle variables $\thpm,\thz$. To do this, we need to evaluate the 
partial derivatives of Hamilton's characteristic function $W$ according to equation (\ref{eq:ang_vars}).
When calculating these derivatives, the coordinates $\xipm$ and $\xiz$ are to be held fixed.
In terms of these coordinates, the contributions to $W$ have the explicit form 
\begin{eqnarray}
W_{+} &=& (2\ip + |\iz|) \Biggl\{
\frac{1}{2}\arccos \left( \frac{\xipmax + \xipmin -2\xip}{\xipmax - \xipmin} \right)
  + \frac{\sqrt{(\xipmax - \xip)(\xip - \xipmin)}}{\xipmax + \xipmin}
\Biggr\}
\nonumber
\\
&&
-|\iz| \arctan \sqrt{\frac{\xip/\xipmin -1}{1 - \xip/\xipmax}} \,,
\end{eqnarray}
%
%
%\begin{eqnarray}
%W_{+} &=& (2\ip + |\iz|) \Biggl\{
%\frac{1}{2}\arccos \left( \frac{\xipmax + \xipmin -2\xip}{\xipmax - \xipmin} \right)
%\nonumber
%\\
%&& \phantom{ (2\ip + |\iz|) \Biggl\{ }
% + \frac{\sqrt{(\xipmax - \xip)(\xip - \xipmin)}}{\xipmax + \xipmin}
%\Biggr\}
%\nonumber
%\\
%&&
%-|\iz| \arctan \sqrt{\frac{\xip/\xipmin -1}{1 - \xip/\xipmax}} \,,
%\end{eqnarray}
%
%
\begin{eqnarray}
W_{-} &=& (2\im + |\iz|) \Biggl\{
\frac{1}{2}\arccos \left( \frac{\ximmax + \ximmin -2\xim}{\ximmax - \ximmin} \right)
  + \frac{\sqrt{(\ximmax - \xim)(\xim - \ximmin)}}{\ximmax + \ximmin}
\Biggr\}
\nonumber
\\
&&
-|\iz| \arctan \sqrt{\frac{\xim/\ximmin -1}{1 - \xim/\ximmax}} \,,
\end{eqnarray}
\begin{equation}
W_{0} = \iz \xiz \,.
\end{equation}
In the above equations, the dependence of the turning points $\xipmmin,\xipmmax$ on the actions 
$\ipm,\iz$ is given explicitly by
\begin{equation}
\xipmax = \frac{
\left(\ip + \im + |\iz|\right)
\left(\sqrt{\ip + |\iz|} + \sqrt{\ip} \right)^{2} 
}
{\mass k} \,,
\end{equation}
\begin{equation}
\xipmin = \frac{
\left(\ip + \im + |\iz|\right)
\left(\sqrt{\ip + |\iz|} - \sqrt{\ip} \right)^{2} 
}
{\mass k} \,,
\end{equation}
\begin{equation}
\ximmax = \frac{
\left(\ip + \im + |\iz|\right)
\left(\sqrt{\im + |\iz|} + \sqrt{\im} \right)^{2} 
}
{\mass k} \,,
\end{equation}
\begin{equation}
\ximmin = \frac{
\left(\ip + \im + |\iz|\right)
\left(\sqrt{\im + |\iz|} - \sqrt{\im} \right)^{2} 
}
{\mass k} \,.
\end{equation}

A long but straightforward calculation now yields
\begin{eqnarray}
\thp &=& \psip - \cp \sin \psip - \cm \sin \psim \,,
\\
\thm &=& \psim - \cp \sin \psip - \cm \sin \psim \,,
\\
\thz &=& \xiz + \frac{\sgn \iz}{2} \left(
\psip + \psim - \chip - \chim  -2c_{+}\sin\psip -2c_{-}\sin\psim
\right) \,,
\end{eqnarray}
where we have defined 
\begin{eqnarray}
\label{eq:cpl}
\cp &=& \frac{\mass k \Ap \epsp}{2I^2} 
= \frac{\sqrt{\ip(\ip + |\iz|)}}{\ip + \im + |\iz|} \,,
\\
\label{eq:cmin}
\cm &=& \frac{\mass k \am \epsm}{2I^2} 
= \frac{\sqrt{\im(\im + |\iz|)}}{\ip + \im + |\iz|} \,.
\end{eqnarray}
%
%
%Note that as $\xip$ runs from $\xipmin$ to $\xipmax$ while $\xim$ is held fixed,
%the value of $\thp$ increases by $\pi$. 
%Similarly, as $\xim$ runs from $\ximmin$ to $\ximmax$ while $\xip$ is held fixed, 
%the value of $\thm$ increases by $\pi$. 
%

In terms of the action-angle variables, the Hamiltonian is given by
(see equation (\ref{eq:ham_az_act})) 
\begin{equation}
\cH = -\frac{\mass k^2}{2(\ip + \im + |\iz|)^2} \,,
\end{equation}
and Hamilton's equations read
\begin{eqnarray}
\label{eq:hameq_aa_1}
\dot\thp = \frac{\partial \cH}{\partial \ip} = \omega \,,
\qquad
\dot\thm = \frac{\partial \cH}{\partial \im} = \omega \,,
\qquad
\dot\thz = \frac{\partial \cH}{\partial \iz} = (\sgn \iz) \omega \,,
\\
\label{eq:hameq_aa_2}
\dot\ip = - \frac{\partial \cH}{\partial \thp} = 0 \,,
\qquad
\dot\im = - \frac{\partial \cH}{\partial \thm} = 0 \,,
\qquad
\dot\iz = - \frac{\partial \cH}{\partial \thz} = 0 \,,
\end{eqnarray}
where
\begin{equation}
\label{eq:eqKeplerf}
\omega = \frac{\mass k^{2}}{(\ip + \im + |\iz|)^{3}} = \frac{\mass k^{2}}{n^3 \hbar^3}
\end{equation}
is the angular Kepler frequency, which only depends on the principal quantum number $n$.
The solution to Hamilton's equations can be immediately
written down as
\begin{equation}
\thp(t) = \omega t + \thpz \,,
\qquad
\thm(t) = \omega t + \thmz \,,
\end{equation}
\begin{equation}
\thz(t) = (\sgn \iz) \omega t + \thzz \,.
\end{equation}
Three integration constants appear in these solutions, but they should not concern us too much in the context of the problem with a weak external field turned on ($F \ne 0$),
as will become clear further below.

\subsubsection{Geometry of the Kepler ellipse}
\label{sec:theory1sub6}

Now, it is well known that when $F=0$ and $E<0$, the motion of the electron
traces out an ellipse. Here we will express the geometrical properties of this
Kepler ellipse --- the semi-major axis $a$, and the eccentricity $\eps$, 
in terms of the constants that arise
in the parabolic-coordinate action-angle formulation of the problem. We 
begin with the two expressions that relate $a$ and $\eps$ to the energy
and angular momentum, which can be found in classical mechanics textbooks presenting the orbit shape of the Kepler problem~\cite{Goldstein}:
\begin{equation}
\label{eq:kepler_spherical}
a = - \frac{k}{2E} \,,
\qquad
\eps = \sqrt{1 + \frac{2EL^2}{\mass k^2}} \,.
\end{equation}
It follows from the first equation of (\ref{eq:apm_epspm}) that
\begin{equation}
\label{eq:semimajor_axis}
a = \frac{\Ap + \am}{2}
= \frac{\xipmin + \xipmax + \ximmin + \ximmax}{4} 
= \frac{(\ip + \im + |\iz|)^2}{\mass k}
=\frac{n^2 \hbar^2}{\mass k}
\,.
\end{equation}

Obtaining the eccentricity $\eps$ takes more work. We begin by constructing two different
linear combinations of the angle variables that are constant in time, named $\delta$ and $\zeta$, which is possible
on account of the doubly-degenerate frequencies (\ref{eq:hameq_aa_1}):
\begin{eqnarray}
\delta &=& \thp - \thm
\nonumber
\\
&=&
\thpz - \thmz
\nonumber
\\
&=& \psip - \psim\,,
\label{eq:delta}
\end{eqnarray}
and
\begin{eqnarray}
\zeta &=& \thz - \frac{1}{2}(\sgn \iz) (\thp + \thm)
\nonumber
\\
&=& \thzz -\frac{1}{2}(\sgn \iz) (\thpz + \thmz)
\nonumber
\\
&=& \xiz -\frac{1}{2}(\sgn \iz)(\chip + \chim) \,.
\label{eq:zeta}
\end{eqnarray}
Now, the solution of the Kepler problem in spherical polar coordinates
makes use of the {\it eccentric anomaly} $\psi$, which has a well-known 
geometric interpretation in terms of the Kepler ellipse, and is related to $r$ by
\begin{equation}
r = a (1 - \eps \cos \psi) \,.
\end{equation}
Using (\ref{eq:eq1}) to express $r$ in terms of $\xipm$, and then 
using the above equation, along with the first equations of
(\ref{eq:psip})-(\ref{eq:psim}), as well as (\ref{eq:semimajor_axis}), yields
the following relationship between $\psi$ and $\psipm$:
\begin{equation}
\eps \cos \psi = \frac{\Ap\epsp\cos\psip + \am\epsm\cos\psim}{\Ap + \am} \,.
\end{equation}
Combining the above with (\ref{eq:delta}), and using some basic trigonometric identities yields
\begin{eqnarray}
\label{eq:eps_delta}
\eps &=& \frac{\sqrt{\Ap^2\epsp^2 + \am^2\epsm^2 +2\Ap\epsp\am\epsm\cos\delta}}{\Ap + \am} 
\\
&=& \sqrt{\cp^2 + \cm^2 +2\cp\cm\cos\delta} \,,
\end{eqnarray}
where the second expression is in terms of dimensionless quantities, given in terms of the actions, cf.~(\ref{eq:cpl},\ref{eq:cmin}).
Thus, the constant phase difference $\delta$ between the two parabolic oscillators is related
to the eccentricity $\eps$ of the Kepler ellipse, or equivalently, via equation (\ref{eq:kepler_spherical}),
to the angular momentum 
magnitude $L$.

Understanding the physical significance of the other constant, namely $\zeta$ (defined in equation (\ref{eq:zeta})) requires
more work, and is not pursued in this paper.

\subsubsection{Solution with perturbing electric field}
\label{sec:theory1sub7}

So far, the discussion of action-angle variables has only been in the context of the unperturbed Kepler problem.
Now we add the perturbing electric field
\begin{equation}
\Delta V = Fz \,.
\end{equation}
An exact solution to the Kepler-Stark problem is obtained by adding the above perturbation to the Hamiltonian $\cH$, 
expressing $z$ in terms of the action-angle variables, and solving the modified Hamilton equations 
(\ref{eq:hameq_aa_1})-(\ref{eq:hameq_aa_2}). In practice, 
this involves the double Fourier series%
\begin{equation}
z(\ip,\im,\iz,\thp,\thm,\thz) =
\sum_{k = -\infty}^{\infty}
\sum_{l = -\infty}^{\infty}
C_{k,l}(\ip,\im,\iz) \, 
e^{i(k\thp + l\thm)} \,.
\end{equation}
Note that $z$ does not depend on $\thz$, since this angle variable is associated with azimuthal motion in the $xy$ plane,
and this is the reason for the series having two rather than three indices. 
In general, when expressing functions of $(x,y,z)$ in terms of the action-angle variables,
one obtains triple Fourier series with phase factors $e^{i(k\thp + l\thm + m\thz)}$.

For weak electric fields
(small $F$), we can obtain
approximate solutions by averaging the perturbation $\Delta V$, which amounts to keeping only the DC part of the Fourier
series, and removing the oscillatory terms:
\begin{equation}
\label{eq:dv_avg}
\Delta V \to F C_{0,0}
=
\frac{3FII_{e}}{2\mass k} \,.
\end{equation}
Upon carrying out this averaging procedure, the solution of Hamilton's equations 
(\ref{eq:hameq_aa_1})-(\ref{eq:hameq_aa_2}) for the perturbed problem becomes a 
straightforward matter. The action variables $(\ip,\im,\iz)$ remain constant,
and the angle variables remain linear functions of time, with perturbed frequencies
\begin{equation}
\label{eq:thdotpm_pert}
\dot{\thp} = \omega + \frac{3F(I+I_{e})}{2\mass k} \,,
\qquad
\dot{\thm} = \omega - \frac{3F(I-I_{e})}{2\mass k} \,,
\end{equation}
\begin{equation}
\label{eq:thdotz_pert}
\dot{\thz} = (\sgn \iz) \left[ \omega + \frac{3FI_{e}}{2 \mass k}  \right] \,.
\end{equation}
This frequency splitting causes the phase difference $\delta$ between
the `plus' and `minus' oscillators (see equation (\ref{eq:delta})) to evolve linearly in time according to
\begin{equation}
\dot{\delta} = \frac{3FI}{\mass k} \,.
\end{equation}
Combining the above with equation (\ref{eq:eps_delta}) yields the evolution of the eccentricity of the ellipse.
Note that the other quantity $\zeta$ (defined in equation (\ref{eq:zeta})) remains constant, even in the presence
of the frequency splitting, since it can be shown from equations~(\ref{eq:thdotpm_pert})-(\ref{eq:thdotz_pert})  that
\begin{equation}
\dot{\zeta} = 0 \,.
\end{equation}
%

%\pagebreak

%

%

\subsection{Quantum mechanical approach}
\label{sec:theory2}

\subsubsection{Connection of radiative Stark-state transition rates to rates for angular momentum eigenstates}
\label{sec:theory2sub1}

The Stark states  labeled by ($n, \mu=n_1-n_2, m$) can be decomposed in terms of the spherical basis states (angular momentum eigenstates) labeled by ($n, l, m$).
Note that we switch notation when comparing the quantum numbers with the semi-classical ones in accord with $n_{+} = n_1$ and $n_{-} = n_2$ to follow the more common 
notation for the remainder of the paper.
The expansion coefficients can be written in terms of a Wigner $3j$ symbol as~\cite{Gallagher_1988}
\begin{equation}
c^{n  \mu m}_{n l m} = \langle n l m| n \mu m \rangle = (-1)^m \sqrt{2l+1} 
\begin{pmatrix}
\frac{1}{2}(n-1)&\frac{1}{2}(n-1) & l \\
\frac{1}{2}(m+\mu)& \frac{1}{2}(m-\mu)& -m
\end{pmatrix} \ ,
\label{eq:Wigner}
\end{equation}
which is a consequence of being able to solve the parabolic oscillators using angular momentum algebra~\cite{Park_1960, Fujio_1977}.
Further studies of the relationship between the quantum states in these representations are found in Refs.~\cite{Tarter_1970, Hey_2007}.
A discussion of features of the lifetimes of $m=0$ vs $m \ne 0$  Stark states was given in Refs.~\cite{PhysRev.133.A424,OMIDVAR1983215},
where an explanation of these features is provided on the basis of symmetries in the matrix elements and sum rules.

Einstein A-coefficients for the parabolic states yielding transition rates from state $(n, \mu, m)$ to state $(n', \mu', m')$ can be expressed as~\cite{PhysRevA.12.1949,Seiler_2016}
\begin{equation}
A_{(n, \mu, m) \to (n', \mu', m')} \sim  \left|  \sum_{l=|m|}^{n-1}  \sum_{l'=|m'|}^{n'-1}   \langle n l m | n \mu m \rangle \langle n' l' m' | n' \mu' m' \rangle
\langle n' l' m' |  e {\hat {\bf r}} | n l m\rangle  \right |^2 \ ,
\end{equation}
where we have used the fact that the expansion coefficients $c^{n  \mu m}_{n l m} = \langle n l m | n \mu m \rangle$ are real-valued.

A number of  selection rules restrict the sums, such as the condition that $\Delta l = \pm 1$ from the dipole operator matrix elements limits the
sums over $l, l'$; the principal quantum numbers for the final states are below the original one  $n' = 1, \dots , n-1$ (we are ignoring weak transitions
that can happen within a multiplet),
and $\Delta m =-1,0,1$ limits the relations between $m, m'$.
One can then define partially summed transition rates with fixed $\Delta n=n-n'$, i.e., for  a given initial state $(n,\mu, m)$ one sums
over allowed $\mu'$ and $m'$, and the overall lifetime for this state follows from a complete sum of A-coefficients and taking the inverse.

Fortunately, there is a shortcut which may play a significant role in allowing the classical-quantum correspondence: there are no interference terms
if one considers decays of states $(n, \mu, m)$~\cite{PhysRevA.12.1949}. One can obtain the partial rates from a procedure that has the following interpretation:
the spontaneous decay rate of a Stark state can be obtained by just looking at the $l$-content of the state and combining the
decay rates with the appropriate weights.

Using the squares of the coefficients $\langle n l m | n \mu m \rangle$ as weights one can then obtain the partial $\Delta n$ dependent transition rates for the
Stark states in terms of the well-known rates for the angular momentum eigenstates.
Assuming nuclear charge $Z=1$, and ignoring reduced-mass effects, i.e., assuming an infinitely heavy nucleus ($m_e$ is the electron mass), 
the rates for transitions
from state $(n,l)$ to $(n',l')$ can be written in terms of a radial matrix element as
\begin{equation}
R_{n,l \to n',l'}=\frac{m_e^3 \alpha^7 c^4 \max{(l,l')}}{6 \hbar^3(2l+1)}\left| \langle n l | r | n' l' \rangle \right|^2 \left(\frac{1}{n'^2} - \frac{1}{n^2} \right)^3 \ ,
\end{equation}
where the last term is related to the phase-space factor, i.e., the cube of the difference in level energies.

For fixed azimuthal quantum number $m$ starting from a state with given $(n, \mu)$ the decay rate to lower principal quantum number shells
(which will be the main concern of this paper) is then calculated as 
\begin{equation}
R_{n,\mu, \Delta n}=\sum_{l=|m|}^{n-1}{\left( c^{n \mu m}_{n l m} \right)^2 (R_{n,l \to n-\Delta n, l-1}+ R_{n,l \to n-\Delta n, l+1}) } \ .
\end{equation}

In order to highlight the classical-quantum correspondence it is useful to look at a few cases of distributions of the weights over the 
angular momentum quantum number $l$.

\subsubsection{Quantum vs semi-classical angular momentum distributions}
\label{sec:theory2sub2}

\begin{figure}
\begin{center}$
%\resizebox{0.6\textwidth}{!}{%
\begin{array}{ccc}
\resizebox{0.5\textwidth}{!}{\includegraphics{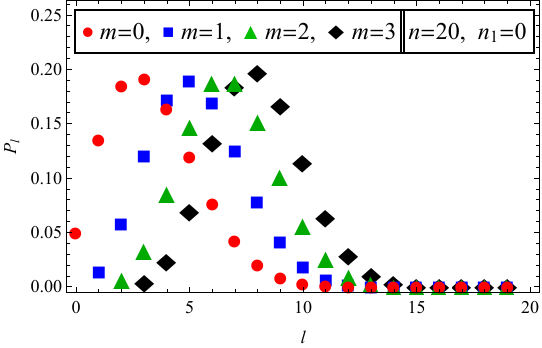}}&
\resizebox{0.5\textwidth}{!}{\includegraphics{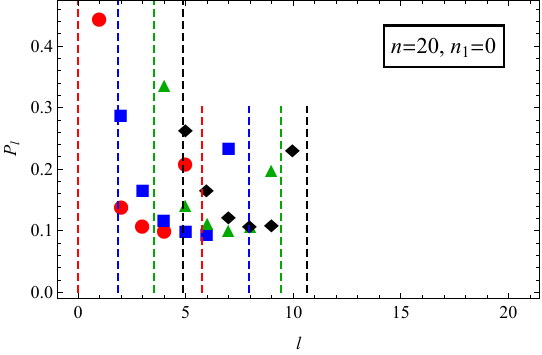}}&\\
\resizebox{0.5\textwidth}{!}{\includegraphics{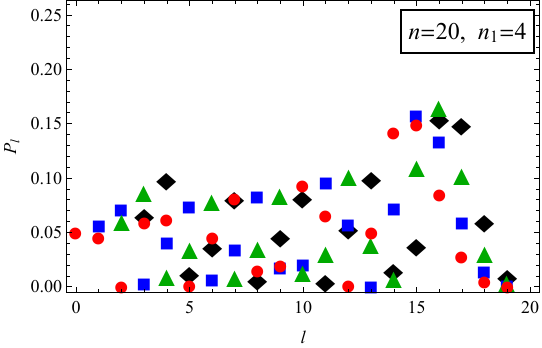}}&
\resizebox{0.5\textwidth}{!}{\includegraphics{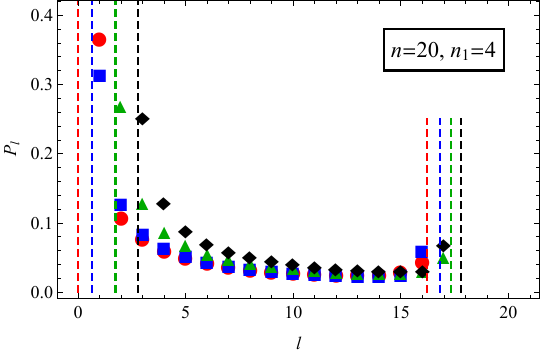}}&\\
\resizebox{0.5\textwidth}{!}{\includegraphics{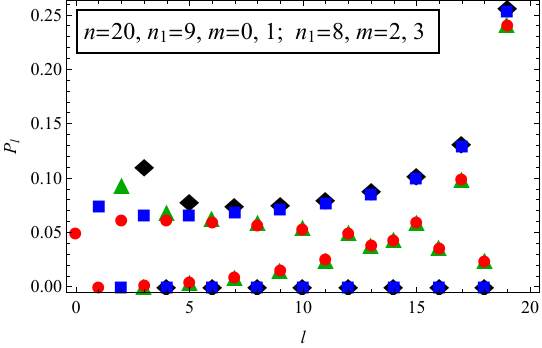}}&
\resizebox{0.5\textwidth}{!}{\includegraphics{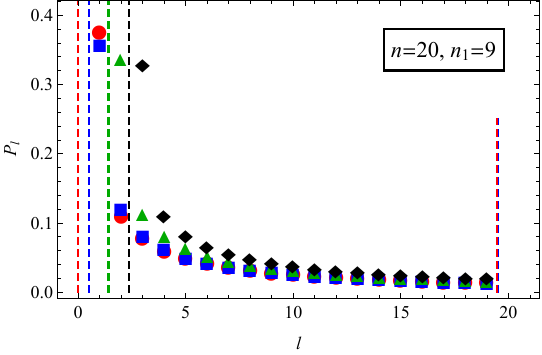}}
\end{array}$
%}
\caption{%
Weights $P_l$ (the squares of coefficients (\ref{eq:Wigner})) for the parabolic states for $n=20$ and $m=0,1,2, 3$ are shown in the left panels, 
while the right panels display corresponding results from integration of classical angular momentum distributions $dP/dl$  over ranges 
$[l-\tfrac{1}{2},l+\tfrac{1}{2}]$.
The left and right boundaries of the classically allowed $l$-ranges are indicated by vertical dashed lines - 
the $dP/dl$ have integrable singularities at these points.
Results are shown for the states  $n_1=0$ (first row), $n_1=4$ (middle row),
and in the bottom row for the central states given by  $n_1=9$ for $m=0,1$, and $n_1=8$ for $m=2,3$, since $n_2=n-m-1-n_1$. 
}
\label{fig:Abb1}
\end{center}
\end{figure}

Fig.~\ref{fig:Abb1} shows that we can aim for better agreement for the extreme parabolic states (larger magnitudes of $\mu=n_1-n_2$) than for 
the near-central states for which quantum mechanics predicts some parity dependence related to selection rules satisfied by the Wigner $3j$ coefficients.

The extreme states ($n_1=0$, or $\mu=n_1-n_2$ maximally negative) are those with maximal electric dipole moment along the $z$-axis, i.e., they have a maximally skewed charge density.  The symmetric counterparts would be the states with $n_2=0$ and maximally positive $\mu$. They arise as a combination of one of the parabolic oscillators being
in a nodeless state. For the semi-classical distributions (which are obtained as functions of $l$ by using the Langer WKB association between eccentricity and angular momentum~(\ref{eq:Langer}))
we find a strong selection of angular momentum ranges which are in accord with the quantum result. There is a concentration of angular momentum values at the boundaries
determined by the classical turning points of the parabolic oscillators. Integration of the continuous classical distribution $dP/dl$ near the inner turning points leads to a large increase in 
probability associated with the smallest integer $l$-value.

For intermediate values of $n_1$ the character of the quantum-mechanical distribution over angular momentum value $l$ changes to an oscillatory pattern.
The suppression on the small-$l$ end is due to the fact that $l \ge |m|$, but there is also $m$-dependent suppression at the high end of the $l$-range, below the sharp
cutoff given by $l=n-1$. This is the result of the main peak in the distributions over $l$ shifting towards the limiting value as $m$ increases.
The quantum mechanical probabilities for the truly central states (which occur for $m=1, 3$) reflect that these Stark states
 are parity eigenstates, i.e., they are superpositions of purely even- or odd-$l$ angular momentum eigenstates.

\subsubsection{Classical Coulomb-Stark orbits}
\label{sec:theory2sub3}

The classical distributions display similar behavior by allowing extended ranges with similar features at both ends of the $l$-range.
The functions $dP/dl$ are, of course, smooth and average through the quantum oscillatory patterns. This trend continues for the 
central states, for which the quantum pattern is one of rapid oscillations with the odd-$m$ cases suppressing even $l$-values.
This follows from a symmetry property of the Wigner $3j$ symbol discussed in Ref.~\cite{Park_1960} for the case of $n_1=n_2$ (or $\mu=0$),
which occurs for odd $m$ when $n$ is even.
For the even-$m$ cases the suppression of odd-$l$ contributions is visible at small $l$ only.
For larger $l$ the probabilities depend mostly on the angular momentum parity only, i.e., probabilities $P_l$ for $m=0,2$ and $m=1,3$ 
are very similar respectively. 

The classical distributions $dP/dl$ again are free of oscillatory behavior, they show the $m$-dependent 
suppression of small-$l$ contributions due to the presence of centrifugal barriers in the parabolic oscillators. Common to both the intermediate
and central values of $n_1$ is that in the quantum case there are large-$l$ admixtures, which apparently are not as dominant in the classical case.

The classical distributions are valid for two principally different situations: in the case of zero external field ($F=0$) one can think of the classical states as ensembles
of ellipses with fixed semimajor axis $a$ determined by the principal quantum number $n$. The distribution of eccentricities $\epsilon$ follows from the demand
to allow all possible relative phases of the parabolic oscillators, i.e., to distribute $\delta$ uniformly, cf. equation (\ref{eq:delta}). 
If one has the situation with a weak electric field present, i.e., $F \ne 0$,
one can start the two parabolic oscillators with any relative phase, and the state will evolve in time such as to visit all possible eccentricities periodically. The motion is one
of ellipses which 'breathe' and precess at the same time. For the extreme states $n_1=0$ (or $n_2=0$) the ensemble of ellipses (or the evolving ellipse)  approach the shape of a paraboloid~(\ref{eq:paraboloid}). They would reside perfectly on a paraboloid if one of the two actions was exactly zero, instead of $\hbar/2$ (cf. equation(\ref{eq:principalaction})).

\begin{figure}
\begin{center}$
%\resizebox{0.6\textwidth}{!}{%
\begin{array}{ccc}
\resizebox{0.5\textwidth}{!}{\includegraphics{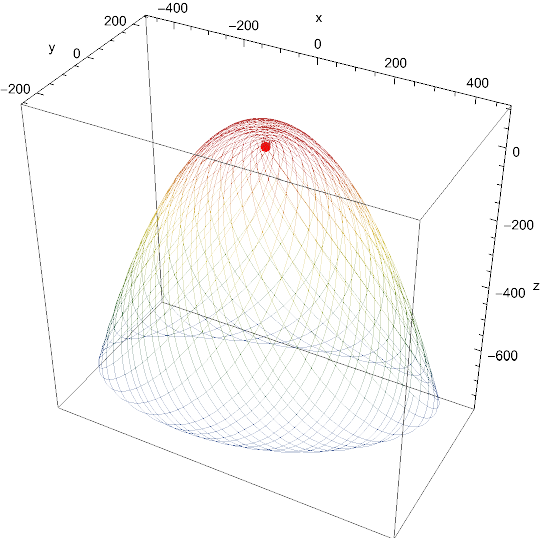}}&
\resizebox{0.5\textwidth}{!}{\includegraphics{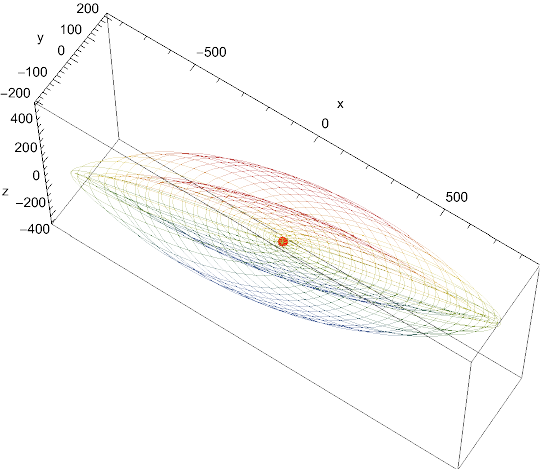}}
\end{array}$
%}
\caption{%
Representation of orbits which are precessing and breathing ellipses due to the presence of a weak electric field along the $z$-axis.
The extreme state $n=20, m=3, n_1=0, \mu=-16$ is shown on the left, and the central state $n=20, m=3, n_1=8, \mu=0$ on the right.
The proton is located at the origin and is represented by a red dot.
See text for further explanations.}
\label{fig:Abb1A}
\end{center}
\end{figure}

This is illustrated in Fig.~\ref{fig:Abb1A} by showing a discrete point plot of locations visited by the classical electron.
For the extreme state $n_1=0, \mu=-16$ (left panel) the points show a view of the paraboloid which encloses the origin and is open towards 
negative $z$-values. A large permanent electric dipole moment emerges due to this asymmetric distribution.

For the central state the distribution over $z$ is nearly symmetric (it would be perfectly symmetric for $\mu=0$ with zero electric dipole moment), and the shape is more difficult
to describe. With increasing $m$-value the ellipses become near-circular as one approaches the limit of the allowed range of values.

\begin{figure}
\begin{center}$
%\resizebox{0.6\textwidth}{!}{%
\begin{array}{ccc}
\resizebox{0.5\textwidth}{!}{\includegraphics{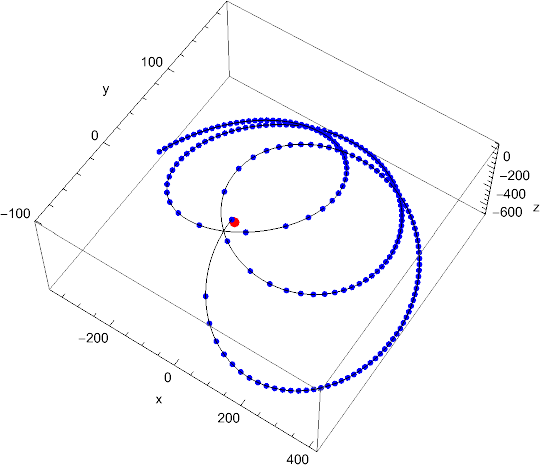}}&
\resizebox{0.5\textwidth}{!}{\includegraphics{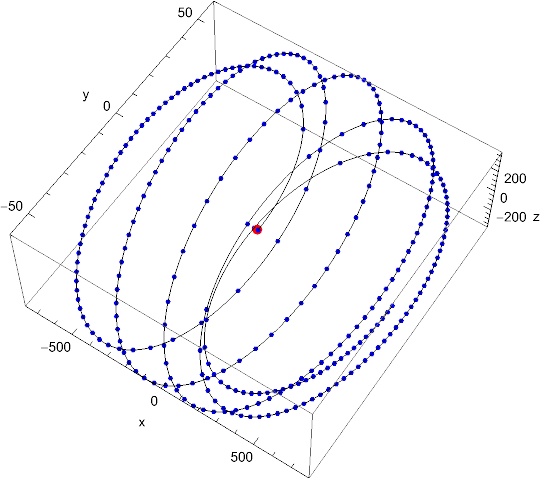}}
\end{array}$
%}
\caption{%
Representation of the early time evolution of orbits which are precessing and breathing ellipses due to the presence of a weak electric field along the $z$-axis.
The extreme state $n=20, m=3, n_1=0, \mu=-16$ is shown on the left, and the central state $n=20, m=3, n_1=8, \mu=0$ on the right.
The orbit is represented by both a continuous black line, and by snapshot locations at constant time intervals $\Delta t$. The electron orbit starts close to the proton,
and eventually covers space as shown in Fig.~\ref{fig:Abb1A}.}
\label{fig:Abb1B}
\end{center}
\end{figure}

In order to illustrate how Larmor radiation arises and varies as a function of the electronic acceleration we illustrate the orbit shapes for the same states in Fig.~\ref{fig:Abb1B}.
In both cases the trajectory starts near the proton. As the electron visits the proton vicinity the acceleration is large, and the contribution from Larmor radiation would be large.
The orbit is calculated without back-reaction, i.e., the problem of radiative loss is treated separately without influence on the electronic orbit.
 
\section{Results}
\label{sec:res}

\subsection{Transition rates for $m=0$ vs $m \ne 0$ states}
\label{sec:result1}

\subsubsection{Lifetimes or total transitions rates by summation over $\Delta n$-transitions}
\label{sec:result1sub1}

\begin{figure}
\begin{center}$
%\resizebox{0.6\textwidth}{!}{%
\begin{array}{ccc}
\resizebox{0.5\textwidth}{!}{\includegraphics{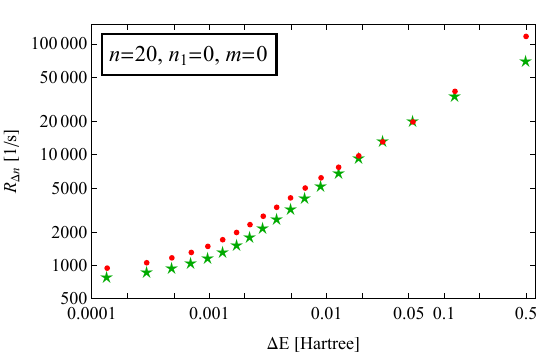}}&
\resizebox{0.5\textwidth}{!}{\includegraphics{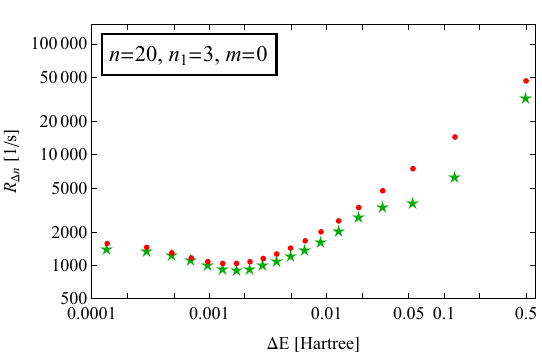}}&\\
\resizebox{0.5\textwidth}{!}{\includegraphics{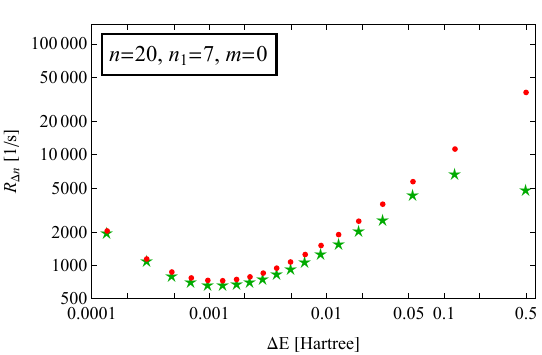}}&
\resizebox{0.5\textwidth}{!}{\includegraphics{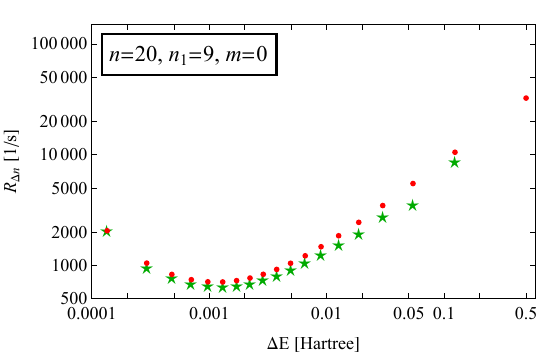}}
\end{array}$
%}
\caption{%
Transition rates to states with $\Delta n =1,2,...$ are shown as a function of radiated energy for the states $n=20$ and $m=0$. 
The four panels show selected examples of initial states starting with the extreme $n_1=0$ state and then stepping towards the central state $n_1=9,n_2=10$.
The quantum rates are shown as stars, the semi-classical rates as circles. For the pair of central states the quantum decay rate to the ground state is about $100 \ \rm s^{-1}$, i.e.
the data point falls below the horizontal axis.
}
\label{fig:Abb2}
\end{center}
\end{figure}

In Fig.~\ref{fig:Abb2} a comparison of classical and quantum transition rates is given for three representative states of the $n=20, m=0$ multiplet.
The circles show the classical rate calculation results that are based on a sampling of the eccentricities that are allowed by the Hamilton-Jacobi
approach for the given choice of semi-classical quantized actions. For each possible ellipse one has a Fourier representation of the orbit which leads
to rate expressions as a function of harmonic order $k$ for both angular-momentum decreasing and increasing transitions (\ref{eq:Clrate}).
The number of harmonics required to cover the energy range down to the ground state is large, on the order of $K \sim 10^4$, since 
 $\omega_0 \approx (E_{20}-E_{19})/\hbar$ and
$K \omega_0 \approx (E_{20}-E_{1})/\hbar$. This calculation would be tedious and one would need to sum the harmonic orders to form energy window bins that
correspond to given $\Delta n$. The re-scaling to non-integer index is given by
\begin{equation}
k_{\Delta n} = \frac{n}{2}\left[( 1-\Delta n /n)^{-2} -1\right] 
\end{equation}
which also implies a weighting of this effective harmonic transition rate by a factor of $( 1-\Delta n /n)^{-3}$, (cf. Eqs.~(22,23) in Ref.~\cite{PhysRevA.71.020501}).

The classical rates for the Rydberg-Stark states in the limit of $F \rightarrow 0$ are obtained by an average of the decay rates for a Kepler ellipse over the
range of eccentricities that follows from the allowed motions for the parabolic oscillators. The average over phases $\delta$ between the parabolic oscillators in $\xip$ and
$\xim$ yields an expression for the $\Delta n$-dependent decay rates for states specified by $(n, \mu,m)$ as
\begin{equation}
R_{n \to n-\Delta n}^{(\mu,m)} = \frac{1}{\pi}\int_{\epsilon_{\rm max}}^{\rm \epsilon_{\rm min}}{f(k_{\Delta n},\epsilon) \left( \frac{d \delta}{d \epsilon} \right) d \epsilon} \ ,
\label{eq:eccaverage}
\end{equation}
where the range of integration for the eccentricity follows directly from equation~(\ref{eq:eps_delta}).
The function to be integrated is the total classical decay rate (cf. Eq.~(13) in Ref.~\cite{PhysRevA.71.020501})
\begin{equation}
\label{eq:Clrate}
f(\tilde k,\epsilon)=\frac{\alpha a^2}{3 c^2}(b_{\tilde k}(\epsilon)^2 + c_{\tilde k}(\epsilon)^2) ({\tilde k} \omega)^3 ( 1-\Delta n /n)^{-3} \quad {\rm with} 
\quad {\tilde k}\equiv  k_{\Delta n} \ ,
\end{equation}
and where $\omega$  (given in (\ref{eq:eqKeplerf})) is the angular frequency of the ensemble of elliptic orbits representing the initial state.

The Fourier coefficients for the time evolution of the  Kepler ellipse orbits evaluated at non-integer $\tilde k$ are given in terms of Bessel functions
and their derivatives as
\begin{equation}
b_{\tilde k}(\epsilon) = \frac{2}{\tilde k} J_{\tilde k}'(\tilde k \epsilon) \quad {\rm and } \quad 
c_{\tilde k}(\epsilon)=\frac{2}{\tilde k} \sqrt{(\epsilon^{-2}-1)}J_{\tilde k}(\tilde k \epsilon) \ .
\end{equation}
The computation of the latter at non-integer values of $\tilde k$ does not represent a problem in modern computing environments.

The comparison of these classical and quantum results shows that for about $\Delta n =1, \ldots, 10$ or more the same pattern can be observed.
The classical rates exceed the quantum rates, and this becomes a problem as $\Delta n$ increases.
A dramatic problem occurs for the pair of central states ($n_1=9, n_2=10$), for which the transition rate to the ground state drops by more than an order of 
magnitude compared to the $n=2$ final state. An indication for such behavior is already apparent for the state $n_1=7$.
The physical origin of this drop can be seen in the bottom left panel of Fig.~\ref{fig:Abb1}. For this state there is a vanishing contribution of 
p-state admixture, i.e., $l=1$ angular momentum states are practically absent. This makes the transition to the ground state
by electric dipole transitions very difficult for those states.

One might speculate why the classical radiation rates cannot follow this trend, but continue to predict strong decays
to the ground state as $n_1$ approaches $n/2$, but there is no simple remedy for the problem. For $m=0$ there is no centrifugal barrier
and eccentricities up to the value of $\epsilon=1$ are permitted in this case.
For the angular momentum eigenstates the relation  (cf. Eq.~(2) in Ref.~\cite{PhysRevA.71.020501})
\begin{equation}
\epsilon=\sqrt{1-\frac{(l+1/2)^2}{n^2}}
\label{eq:Langer}
\end{equation}
indicates that s-states are represented by ellipses with $\epsilon<1$. Nevertheless, one shies away from decay rate calculations for $l=0$,
since one would have to impose that angular-momentum-decreasing ($\Delta l =-1$) transitions are not allowed.

Ironically, in the case of the Stark states it is the extreme states which are represented by an ensemble of ellipses with large
eccentricities, and the transition rates follow the pattern of the quantum ones reasonably well - although with a factor-of-two overestimate
of the $\Delta n = n-1$ rate, i.e., for direct transitions to the ground state.

For the near-central states the ensemble covers a wide range of eccentricities. This leads at the same time to very good agreement
for many values of $\Delta n$, but also to a complete breakdown for $\Delta n = n-1$ transitions. This has repercussions for the
overall decay rates of these states, i.e., the pattern of lifetimes of the states as a function of $(n_1, n_2)$ will be affected by the
inclusion of the predicted strong transition rate to the ground state.

To summarize the comparison between the classical and the quantum calculations we can make the following statements. 
For the extreme state $n_1=0$ there is good agreement over a range of $\Delta n$ values, which is reasonable given
the distribution of the quantum state over orbital angular momentum $l$ (Fig.~\ref{fig:Abb1}). Even though the distribution of 
probabilities $P_l$ is remarkably different, the classical ones reflect a reasonable average of the the dominant quantum
mechanical angular momentum states.

 For the central state
the classical result actually works better than one would have expected on the basis of Fig.~\ref{fig:Abb1}. What appears to save the 
classical calculation in this case is the strong admixture of the maximal $l$-value for which radiative transitions are slow and 
 follow along the maximum angular momentum path $l=n-1$ with $\Delta n= 1$ and $\Delta l=-1$ transitions.

\begin{figure}
\begin{center}$
%\resizebox{0.6\textwidth}{!}{%
\begin{array}{ccc}
\resizebox{0.5\textwidth}{!}{\includegraphics{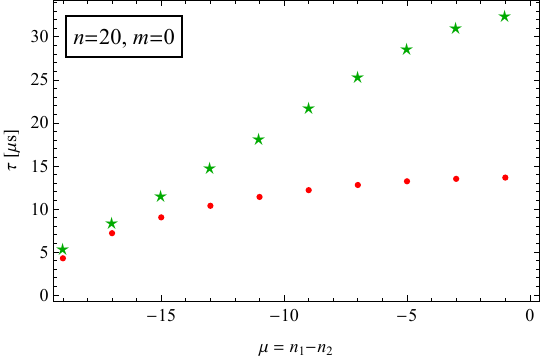}}&
\resizebox{0.5\textwidth}{!}{\includegraphics{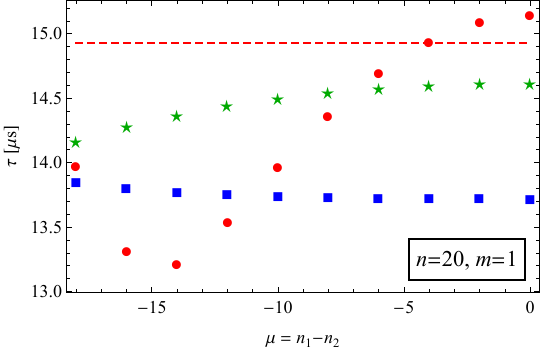}}&\\
\resizebox{0.5\textwidth}{!}{\includegraphics{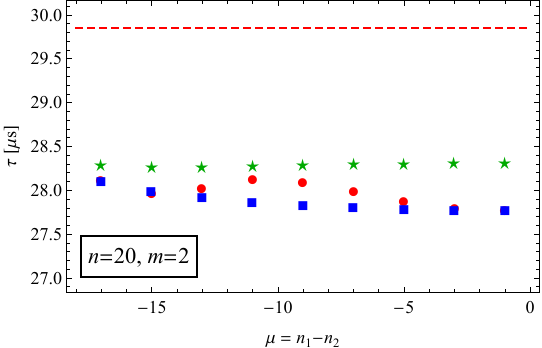}}&
\resizebox{0.5\textwidth}{!}{\includegraphics{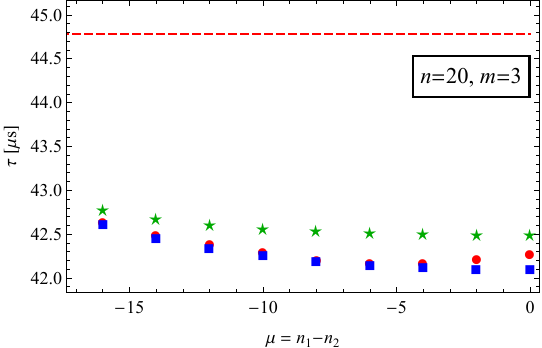}}
\end{array}$
%}
\caption{%
Comparison of classical vs quantum lifetimes  for the Rydberg-Stark levels for $n=20$ and $m=0..3$. 
The classical lifetimes (in microseconds) obtained from a summation of the $\Delta n=1,..n-1$ transitions are shown as disks, the quantum lifetimes
as stars.
For $m=1..3$ the simple-average formula~(\ref{eq:simple}) results are shown as dashed red lines. The points displayed as squares are the result of averaging the more
precise decay rate formula (\ref{eq:Hessels}) that was shown to approximate closely the quantum transition rates for $(n,l)$ states~\cite{Horbatsch_2005}.
}
\label{fig:Abb3}
\end{center}
\end{figure}

The semi-classical lifetimes for the $m=0$ states will be compromised for the reasons mentioned above, but a comparison with the quantum results
is nevertheless provided in Fig.~\ref{fig:Abb3}. The Larmor radiation result is acceptable for the extreme states,
but as one moves towards the central states one notes a rapid deterioration: while the quantum lifetimes increase by a factor of six over the range
of $\mu=n_1-n_2$, the classical lifetimes flatten and fall short by about a factor of two for the central parabolic states. This can be traced mostly to the 
absence of $l=1$ states in the quantum superposition (cf. Fig.~\ref{fig:Abb1}).

\begin{figure}
\begin{center}$
%\resizebox{0.6\textwidth}{!}{%
\begin{array}{ccc}
\resizebox{0.5\textwidth}{!}{\includegraphics{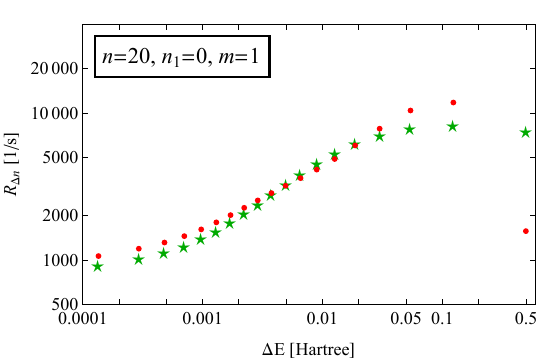}}&
\resizebox{0.5\textwidth}{!}{\includegraphics{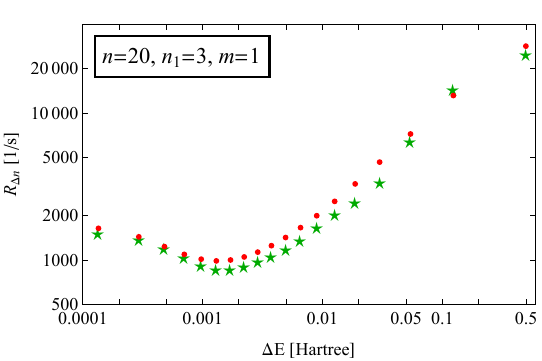}}&\\
\resizebox{0.5\textwidth}{!}{\includegraphics{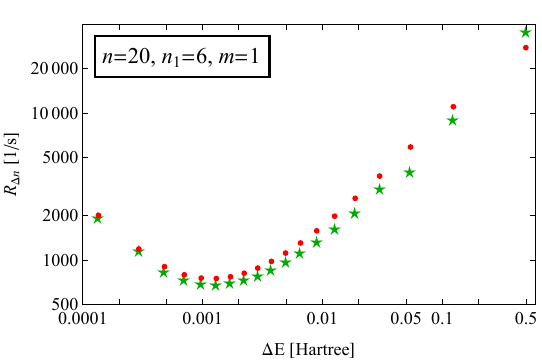}}&
\resizebox{0.5\textwidth}{!}{\includegraphics{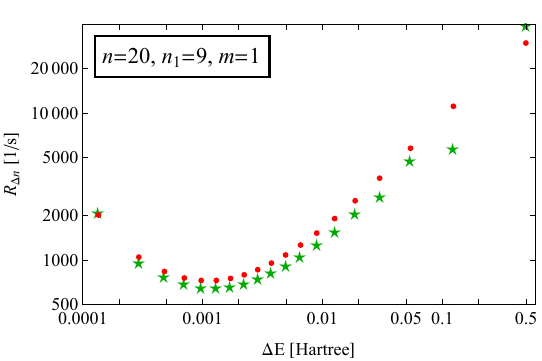}}
\end{array}$
%}
\caption{%
Transition rates  (number of decays per second) to states with $\Delta n =1,2,...$ are shown as a function of radiated energy for the states $n=20$ and $m=1$. 
The four panels show selected examples of initial states starting with the extreme $n_1=0$ state and then stepping towards the central state $n_1=n_2=9$.
The quantum rates are shown as stars, the semi-classical rates as circles.}
\label{fig:Abb4}
\end{center}
\end{figure}

For $m \ne 0$ states we expect better performance from the classical decay calculation, since the inner turning points of the parabolic oscillators are unproblematic.
The detailed $\Delta n$-dependent rates are shown in Fig.~\ref{fig:Abb4} 
for four states spanning the range from the extreme state (with largest electric dipole moment) to the central state.
The classical and quantum rates follow a very similar pattern despite the unusual $l$-distribution.
The practical absence of even-$l$ contributions has no apparent repercussions for the comparison with the Larmor decay rate.
We note, however, a factor-of-three disagreement for transitions to the
ground state for the extreme state $n_1=0$. Here the quantum distribution over angular momentum includes a p-state admixture which decays strongly to
the ground state. The semi-classical distribution over angular momentum starts, however at $l \approx 2$, and this is responsible for the strong drop in 
the electric dipole transition rate to the $n=1$ final state.

Not obvious from  Fig.~\ref{fig:Abb4} is the fact that if one sums the rates over the different $\Delta n$ transitions to calculate the total spontaneous decay rates 
or their inverse, i.e., the lifetimes of the different $n=20, m=1$ but ($n_1,n_2$)-dependent states one obtains approximately the same answer.
This approximate result is shown in the right panel of Fig.~\ref{fig:Abb3} alongside the quantum result, which shows less variation with $\mu=n_1-n_2$.
The other result indicated by a flat line is the result of a calculation which leads to a simple analytic formula for $m \ne 0$ states. This result is obtained by averaging
over eccentricities $\epsilon$ a decay rate formula that is known to give results accurate to about $10 \ \%$ (cf. Eq.~(17) in Ref.~\cite{PhysRevA.71.020501}).
The result of this averaging leads to the expression~\cite{2006APS..DMP.G1021H}
\begin{equation}
\tau \approx \tau_{n,|m|} = \frac{\tau_0}{n |m|} \quad {\rm if} \quad m \ne 0 \ ,
\label{eq:simple}
\end{equation}
which is independent of $\mu=n_1-n_2$, and
the time scale for hydrogenic lifetimes of $n$-levels  (for $Z=1$ and ignoring reduced-mass effects) is given by
\begin{equation}
\tau_0=\frac{3 \hbar n^5}{2\alpha^5 m_{\rm e} c^2} \ .
\end{equation}

A more accurate decay rate formula for angular momentum eigenstates with $l \ge 1$ derived from quantum-classical
correspondence was presented in Eq.~(28) of Ref.~\cite{Horbatsch_2005}:
\begin{equation}
R^{\rm cl}=\frac{1+\frac{19}{88}[(\frac{1}{\epsilon^2}-1)\ln{(1-\epsilon^2)}+1-\frac{\epsilon^2}{2}-\frac{\epsilon^4}{40}]}{\tau_0 (1-\epsilon^2)} \ .
\label{eq:Hessels}
\end{equation}
%For accuracy this expression should be used with a modified $\epsilon-l$ relationship which reads
%\begin{equation}
%\epsilon=\sqrt{1-\frac{l(l+1)+\frac{8}{47} -\frac{l+1}{69 n}}{n^2}} \ .
%\label{eq:epsHessels}
%\end{equation}

Averaging this formula over eccentricities $\epsilon$ in analogy to the calculations performed in Eq.~(\ref{eq:eccaverage}) leads to $n_1$-dependent
lifetime results. This calculation (which is not defined for the case of $m=0$ due to the  $l \ge 1$ restriction) was carried out numerically, and the results are shown in Fig.~\ref{fig:Abb3}
as blue squares. It can be seen that the two accurate classical procedures do not lead to the same answers. In one case we calculated transition rates
as a function of $\Delta n$ by averaging over $\epsilon$ and then summed the rates for all $\Delta n$ (red disks); in the other case (blue squares) 
we used directly an $(n,l)$-dependent decay rate formula that was based on complete summations of the Fourier series for $k=1,2,...$ (in principle to infinity, in practice to reach
numerical convergence) and a fitting procedure to arrive at a formula. The disagreement between the two procedures becomes less significant as one
increases $m$.

\subsection{Quantum-classical correspondence}
\label{sec:result2}

\subsubsection{Correspondence by eccentricity}
\label{sec:result2sub1}

One may ask why the classical calculation works so well in general, particularly when $|m|$ is large. Here we would like to address the issue and also
point out some shortcomings of an approach that does not include motion beyond the classical turning points (a subject matter that can be addressed within
the WKB approach).

\begin{figure}
\begin{center}$
%\resizebox{0.6\textwidth}{!}{%
\begin{array}{ccc}
\resizebox{0.5\textwidth}{!}{\includegraphics{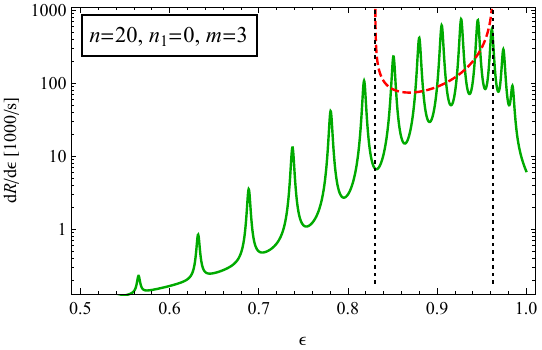}}&
\resizebox{0.5\textwidth}{!}{\includegraphics{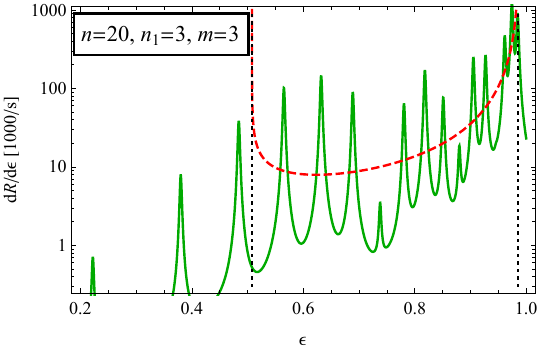}}&\\
\resizebox{0.5\textwidth}{!}{\includegraphics{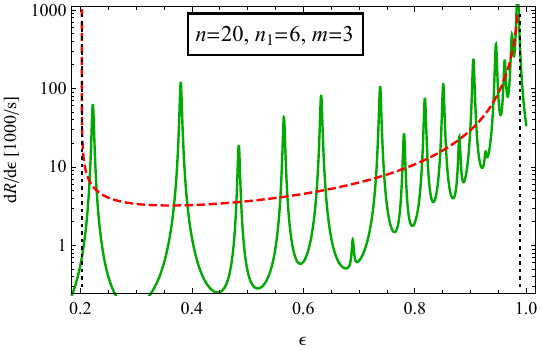}}&
\resizebox{0.5\textwidth}{!}{\includegraphics{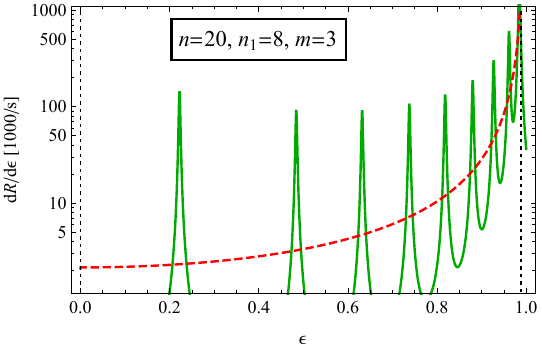}}
\end{array}$
%}
\caption{%
Transition rate as a function of eccentricity $\epsilon$. The classical result is shown as a dashed line with the vertical dotted lines showing the limits given
by $\epsilon_{\rm min}$ and $\epsilon_{\rm max}$. The solid line shows the contributions from the quantum $l$-states expressed in terms of $\epsilon$
by Eq.~(\ref{eq:Langer}), and broadened artificially to show the discrete values as a continuous probability.}
\label{fig:Abb5}
\end{center}
\end{figure}

 In Fig.~\ref{fig:Abb5} the contributions to the classical decay rate are shown to come from a continuous range of eccentricities, and it is obvious that
 for an $m \ne 0$ state the area bounded by this curve and the limits $\epsilon_{\rm min}$ and $\epsilon_{\rm max}$ is finite. The quantum counterpart
 is obtained by first using the weights $P_l$ for the Rydberg-Stark state (given as the squares of  Eq.~(\ref{eq:Wigner}) and shown for some examples
 in Fig.~\ref{fig:Abb1}), associating an eccentricity $\epsilon(n,l)$ according to Eq.~(\ref{eq:Langer}), and then assigning a continuous probability
 distribution by applying a finite-width delta distribution. This function $dP/d\epsilon$ can be multiplied by the transition rate formula given in  
 Eq.~(\ref{eq:Hessels}).
 The result is a pseudo-distribution with $n-|m|-1$ peaks, a few of which are strongly suppressed. The area under this curve yields the quantum decay rate
 which is within one percent of the classical result. Instead of Eq.~(\ref{eq:Langer}) one could also use a more precise association $\epsilon(n,l)$, 
 given as Eq.~(29) in Ref.~\cite{Horbatsch_2005}.
 
The comparison shows that overall the classical result performs some average and that the total decay rate is dominated by the states associated with
the smallest allowed $l$-values ($l=|m|, |m|+1$) and that this occurs for large eccentricities in the vicinity of $\epsilon_{\rm max}$.
At the other end there is the interesting observation that the classical distribution cuts out at $\epsilon_{\rm min}$, while the quantum
pseudo-distribution shows peaks at smaller $\epsilon$. This weakness must have its origin in the fact that the
quantum parabolic oscillator eigenfunctions do have tails outside the classical turning points $\xip^{\rm max}$ and $\xim^{\rm max}$.

\subsubsection{Correspondence of parabolic oscillators}
\label{sec:result2sub2}

\begin{figure}
\begin{center}$
%\resizebox{0.6\textwidth}{!}{%
\begin{array}{ccc}
\resizebox{0.5\textwidth}{!}{\includegraphics{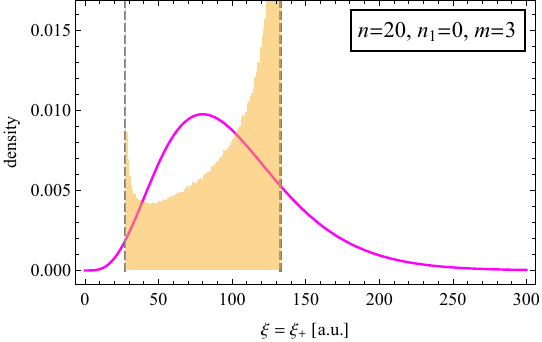}}&
\resizebox{0.5\textwidth}{!}{\includegraphics{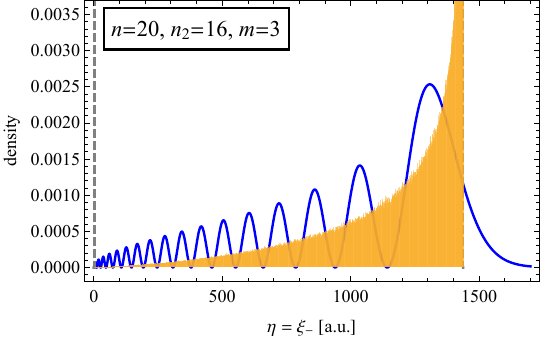}}&\\
\resizebox{0.5\textwidth}{!}{\includegraphics{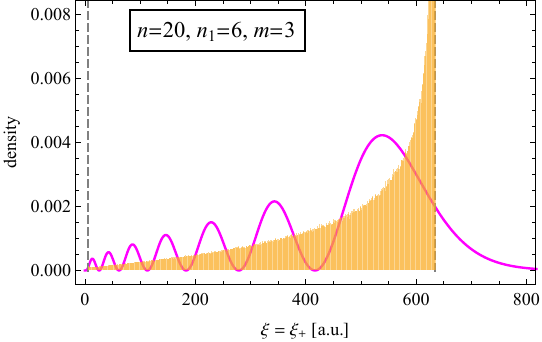}}&
\resizebox{0.5\textwidth}{!}{\includegraphics{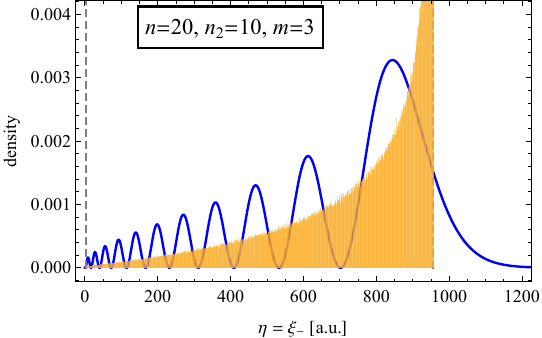}}
\end{array}$
%}
\caption{%
Quantum probability densities for the parabolic oscillator states that combine to create the $n=20, m=3, n_1=0, n_2=16$ Stark state are shown in the top row,
while the bottom row shows the same for the $n=20, m=3, n_1=6, n_2=10$ state. The densities are defined using the solutions given in Ref.~\cite{Hey_2007}.
The classical distributions were generated from histograms for the two coupled oscillators for which the phase $\delta$ was varied uniformly and weights
according to the values of $\xipm$ were applied to the data for a meaningful comparison.
The dashed vertical lines indicate the positions of the classical inner and outer  turning points. All inner turning points occur at non-zero values of the coordinate.
}
\label{fig:Abb6}
\end{center}
\end{figure}

To illustrate further the quantum-classical correspondence for the Rydberg-Stark states we provide in Fig.~\ref{fig:Abb6} density plots for two of the states discussed in this section,
namely the extreme state $n=20, m=3, n_1=0, n_2=16$, and also a state closer to the central state. In  Ref.~\cite{Hey_2007} solutions for the separate
oscillators are provided, and an interpretation can be assigned to them on the basis of Eqs.~(16 d,e) in that work. One can view the result as the probability 
distribution one would obtain if measuring just one of the two coordinates of the electron
(in the notation of Ref.~\cite{Hey_2007} our parabolic coordinate pair is defined as $\xi \equiv \xip$
and $\eta \equiv \xim$).

The question of proper visualization of Rydberg-Stark states has been raised~\cite{kocbach2012visualization} and it concerns the appropriateness
to include (or not) factors that are associated with the volume element. The quantum oscillator wavefunctions that allow a simple interpretation
of one-dimensional motion in an effective potential include a factor $\sqrt{\xipm}$ respectively such that the kinetic energy becomes a simple
second-derivative term. Thus, the quantum densities shown are analogs of radial probabilities, their values yield answers to the question of 
finding the particle within $[\xip, \xip+d\xip )$ and $[\xim, \xim+d\xim )$ respectively during independent measurements.

We note that the (coupled) classical oscillators display outer turning points  in the vicinity where the radial parabolic oscillator orbitals 
(i.e., the functions given in Eqs.(13, 14) of Ref.~\cite{Hey_2007} multiplied by $\sqrt{\xi}$ or $\sqrt{\eta}$ respectively) have inflection points.
The agreement of those locations (quantum vs classical) is very good (three significant digits or better).

For the nodeless 
$n_1=0$ state the inner turning point at the centrifugal barrier is also clearly visible. Here the location of the inflection point in the radial orbital
occurs at  an approximately $10  \%$ smaller value of the coordinate. This is expected since the effective potential in the radial Schr\"odinger equations
(cf. Eqs.~(1,2) in Ref.~\cite{PhysRevA.26.1775}) is different in the centrifugal term on account of the WKB transformation, i.e., this potential is proportional to
$m^2-1$ in quantum mechanics vs $m^2$ in the semi-classical treatment. For the chosen case of $m=3$ this difference is modest, and becomes less
important with increasing $|m|$.

\section{Conclusions}
\label{sec:conclusions}

A comprehensive treatment of the semi-classical Stark problem was presented in parabolic coordinates using the Hamilton-Jacobi formalism.
For states with specified quantum numbers $n$, $m$, and $\mu=n_1-n_2$ the parabolic oscillator solutions allow to define
elliptic orbits with fixed semi-major axis (dependent on the $n$ value), and a given range of eccentricities for fixed $\mu$ and $m$. 
The radiative decays are estimated classically on the basis of the Larmor power by borrowing from previous work for the 
Kepler-Coulomb problem. Decay rates for given $\Delta n$ transitions were obtained and compared to quantum results. For $|m|>0$ states
the results compare well and translate into reasonable values for the overall lifetimes of the states. 

In the case of $m=0$ a careful treatment is required.
Summation of the entire Fourier series (up to infinite-order terms) 
leads to failure in predicting finite transition rates due to the divergence at eccentricity value $\epsilon=1$.
The re-scaled approach, however, in which one collects Fourier terms and associates them with transitions $\Delta n=1,..,n-1$
(cf. Ref.~\cite{PhysRevA.71.020501})
results in finite transition rates and non-zero lifetimes for the $m=0$ Stark states. 
These lifetimes are reasonable for the extreme states, but deviate by a factor of two for the central states $n_1 \approx n_2$
(as demonstrated for the $n=20$ manifold). 
Nevertheless, the classical calculation shares the general feature with the quantum calculation
that for $m=0$ initial states the lifetimes depend strongly on the quantum number $\mu=n_1-n_2$, 
while very little dependence on $\mu$ occurs for all $m>0$ states.

The work complements previous results for decays $(n,l) \to (n', l')$, which were independent of the orientation of the ellipse with known
eccentricity $\epsilon$. It is of interest to note that while the result for a Stark state demands a known value of $m$, the results presented
in this work are independent of the final-state quantum numbers beyond $n'=n-\Delta n$. They were simply derived from the 
properties of the time-dependent motion along the allowed ellipses (in the limit of weak electric field strength $F$ the orbital motion remains bounded).
Future work will involve more detailed branching ratios and will be based on Fourier representations of the three-dimensional orbits which
will allow to distinguish between $\Delta m=-1, 0, 1$ transitions and also consider final distributions over $\mu'=n_1'-n_2'$.

\begin{acknowledgments}
Financial support from the Natural Sciences and Engineering Research Council of Canada (NSERC) is acknowledged. 
We gratefully acknowledge recent discussions with Michael Littman, and also previous discussions with Eric Hessels with whom this project was started some years ago.
We also thank referees for useful suggestions.
\end{acknowledgments}

%
% BibTeX users please use
%\bibliographystyle{plain}
%\bibliographystyle{unsrt}

\bibliography{StarkProblem}

\end{document}